\documentclass[universe,review,accept,oneauthor]{Definitions/mdpi}
\usepackage{amsmath}
\usepackage{amsfonts}
\usepackage{amssymb,bm}
\usepackage{siunitx}
\usepackage{color}

\firstpage{1}
\pubvolume{7}
\issuenum{4}
\articlenumber{84}
\pubyear{2021}
\copyrightyear{2021}
\history{Received: 2 March 2021; Accepted: 23 March 2021; Published: 1 April 2021}

\Title{Casimir Puzzle and Casimir Conundrum: Discovery and Search
for Resolution }

\Author{Vladimir M. Mostepanenko $^{1,2,3}$
}

\address{%
$^{1}$ \quad Central Astronomical Observatory at Pulkovo of the Russian Academy of Sciences, Saint Petersburg, 196140, Russia\\
$^{2}$ \quad Institute of Physics, Nanotechnology and
Telecommunications, Peter the Great Saint Petersburg
Polytechnic University, Saint Petersburg, 195251,  Russia\\
$^{3}$\quad  Kazan Federal University, Kazan, 420008, Russia}

\corres{Correspondence: vmostepa@gmail.com}

\abstract{We review complicated problems in the Lifshitz theory describing
the Casimir force between real material plates made of metals and
dielectrics including different approaches to their resolution.
It has been shown that both for metallic plates with perfect crystal
lattices and for any dielectric plates the Casimir entropy calculated
in the framework of the Lifshitz theory violates the Nernst heat
theorem when the well approved dielectric functions are used in
computations. The respective theoretical Casimir forces are excluded
by the measurement data of numerous precision experiments. In the
literature this situation received the names of the Casimir puzzle
and the Casimir conundrum for the cases of metallic and dielectric
plates, respectively. The review presents a summary of the
main facts on this subject on both theoretical and experimental
sides. Next, we discuss the main approaches proposed in the literature
in order to bring the Lifshitz theory in agreement with the
measurement data and with the laws of thermodynamics. Special attention
is paid to the recently suggested spatially nonlocal Drude-like
response functions which take into account the relaxation properties
of conduction electrons, as does the standard Drude model, but lead
to the theoretical results in agreement with both thermodynamics and
the measurement data through the alternative response to quantum
fluctuations off the mass shell. Further advances and trends in this
field of research are discussed.}


\begin{document}
\newcommand{\ve}{{\varepsilon}}
\newcommand{\kb}{{k_{\bot}}}
\newcommand{\ix}{{{\rm i}\xi}}

\section{Introduction}

The Casimir effect is one of the counterintuitive physical phenomena.
It is somewhat against expectations that two parallel, electrically
neutral ideal metal plates at zero temperature spaced in vacuum at a
distance $a$ attract each other. However, Casimir \cite{1} has
proven that there is an attractive force per unit area of the plates
in this situation which depends on $a$, the Planck constant $\hbar$,
and the speed of light $c$
\begin{equation}
F_{IM}(a)=-\frac{\pi^2}{240}\,\frac{\hbar c}{a^4}.
\label{eq1}
\end{equation}
\noindent
The Casimir force arises due to the zero-point oscillations of quantized
electromagnetic field. In free space the energy density of these
oscillations is infinitely large and unobservable. All the physical
energies are measured from this infinity. In the presence of parallel
plates, however, the tangential component of the electric field and
the normal component of the magnetic induction vanish on their surfaces
and the third component of the wave vector perpendicular to the plates
becomes discrete. In so doing the vacuum energy density in the presence
of plates remains infinitely large but if we, following the common
procedure, subtract from it the vacuum energy density in free space,
the result turns out to be finite
\begin{equation}
E_{IM}(a)=-\frac{\pi^2}{720}\,\frac{\hbar c}{a^3}.
\label{eq2}
\end{equation}
\noindent
This is the Casimir energy per unit area of the plates. In that case
(\ref{eq1}) follows from (\ref{eq2}) by the negative differentiation
with respect to $a$.

Thus, the Casimir effect is directly connected with the concept of the
quantum vacuum which is the most fundamental and not clearly understood
type of physical reality. In spite of the  elaborated in detail
renormalization and regularization procedures, which help to obtain
physically meaningful calculation results, the enormously large vacuum
energy obtained after the momentum cut off is sometimes considered as
catastrophic \cite{2}. It should be mentioned also that the energy of
the quantum vacuum is directly connected with the cosmological
constant \cite{3}, dark energy and the acceleration of the Universe
expansion \cite{4}. This puts the Casimir effect in the list of
closely connected greatest problems of modern physics. It is not
surprising then that experimental and theoretical investigations of
the Casimir force met with outstanding problems which received the names
of {\it the Casimir puzzle} and {\it the Casimir conundrum}.

At first, no fundamental problems in the theory of the Casimir effect
were anticipated. Lifshitz \cite{5,6} developed the general theory
expressing the Casimir free energy and force between two thick plates
(semispaces) kept at any temperature $T$ and made of real materials
described by the frequency-dependent dielectric permittivities. This
theory was generalized for the plates possessing magnetic properties
described by the frequency-dependent magnetic permeability \cite{7}.
It was shown that the familiar van der Waals force is nothing but the
Casimir force at separations between the plates below a few nanometers
where the speed of light can be considered as infinitely large. The
Casimir results (\ref{eq1}) and (\ref{eq2}) were obtained from
the Lifshitz theory at
$T=0$ for the plates made of a perfectly reflecting material. The
Lifshitz results for two thick plates were generalized for the case of
planar systems containing any number of material layers with different
dielectric permittivities \cite{8}.

First experiments on measuring the Casimir force were not enough
precise and have been found only in qualitative agreement with
theoretical predictions of the Lifshitz theory (see \cite{9} for a
review). It was shown \cite{9a}, however, that at large separations
the thermal Casimir force between metallic plates calculated using
the Lifshitz theory and the dielectric permittivity of the
dissipative Drude model is equal to one half of that found for ideal
metal plates at nonzero temperature. This result generated a serious
incomprehension. On the one hand, at low frequencies, which determine
the force behavior at large separations, the Drude model was well
tested in numerous experiments and even in technical applications.
On the other hand, it was expected that with increasing separation
the plate metal should behave more and more closer to the ideal one.
That is why the special prescription was proposed \cite{9a} which is
supplementary to the formalism of the Lifshitz theory (see below).

Wide public attention to the internal problems of this theory was
attracted in 2000 after a publication of the following result. It
was found \cite{10} that at short separations the thermal correction
to the Casimir force computed using the Drude model is relatively
large and decreases the force magnitude, i.e., corresponds to a
repulsion which is quite unexpected. By contrast, the Lifshitz
theory combined with the dielectric permittivity of the
dissipationless plasma model (which should not be applicable at
low frequencies) predicted a smooth approach of the Casimir force
to the ideal metal value at large separations and a very small
thermal correction at short separation. The latter is of the same
sign as the zero-temperature force and only slightly increases its
magnitude \cite{11}.

These problems were dramatized by an inconsistency of the Lifshitz
theory using the Drude model with the laws of thermodynamics. It was
proven \cite{12,13,14} that in the configuration of two metallic
plates with perfect crystal lattices described by the Drude model
the Casimir entropy goes with vanishing temperature to a nonzero
negative constant depending on the parameters of a system in
violation of the third law of thermodynamics (the Nernst heat
theorem). It was also shown \cite{12,13,14} that the Lifshitz theory
using the plasma model leads to the positive Casimir entropy which
goes to zero when the temperature vanishes in agreement with the
Nernst heat theorem. More recently, the same results were obtained
for several other Casimir configurations \cite{15,16,17,18}.

Precise measurements of the Casimir interaction between metallic
test bodies added to the complexity of this situation. In a series
of experiments performed by means of a micromechanical torsional
oscillator and an atomic force microscope the Lifshitz theory using
the Drude model at low frequencies was unambiguously excluded by
the measurement data whereas the same theory using the plasma model
was found to be in good agreement with the measurement results
\cite{19,20,21,22,23,24,25,26}. The dielectric permittivity of a
metal in these experiments was found from the measured optical data
for its complex index of refraction extrapolated by means of either
the Drude or the plasma model to the region of low frequencies
where the optical data are not available. Note that in the
single experiment which was interpreted in favor of the Drude
model \cite{27} the Casimir force was not directly measured
but extracted using the fitting procedure from up to an order
of magnitude larger force presumably determined by the surface
patches. It was shown \cite{28,29} that the interpretation of
this experiment suffers from serious uncertainties.

The situation with dielectric test bodies is closely parallel to
that with metallic ones. It was found that the Casimir entropy
calculated using the Lifshitz theory for ideal dielectrics
(insulators possessing zero electrical conductivity) satisfies the
Nernst heat theorem \cite{30,31,32,33,34,35}. However, at any
nonzero temperature the dielectric permittivity of real dielectric
bodies includes the term describing small but nonzero electric
conductivity at a constant current (the so-called dc conductivity).
If this term is taken into account, as it should be done, the
Casimir entropy calculated using the Lifshitz theory goes to a
nonzero positive constant depending on the parameters of a system
with vanishing temperature, i.e., violates the Nernst heat theorem
\cite{30,31,32,33,34,35}.

The experimental results obtained for dielectric test bodies also
resemble the case of metallic plates. If the dc conductivity of
plate material is taken into account, the predictions of the
Lifshitz theory are excluded by the measurement data \cite{36,37,
38,39,40}. If, however, the dc conductivity is simply omitted in
calculations, the theoretical results are found in agreement with
the data \cite{36,37,38,39,40,41}. Once again, as in the case of
metals, the thermodynamically consistent theory agrees with the
experimental results. This is reached, however, by disregarding
the real physical phenomena -- the dissipation of conduction
electrons for metals and the dc conductivity for dielectrics.
That is why the above problems have been called in the literature
{\it the Casimir puzzle} and {\it the Casimir conundrum}
\cite{42,43,44,45,46}. Taking into account that for metals and
dielectrics the above inconsistencies originate from different
sources (see below), it was suggested \cite{42,46} to call them
a puzzle and a conundrum, respectively.

In this article, we review different efforts to understand both
the puzzle and the conundrum in Casimir physics undertaken in the
literature during the last 20 years. We start with the more
rigorous mathematical formulation of the essence of these problems
and elucidate in more detail the theoretical and experimental
parts of the Casimir puzzle for metallic and the Casimir conundrum
for dielectric test bodies, respectively. The basically viable
approaches to the resolution of these problems can be of two types.
The approaches of the first type search for some physical effects
which could make an impact on the measurement results but were
not taken into account (or were accounted for improperly) in the
comparison between experiment and theory. Among them one could
mention the role of impurities of a crystal lattice and respective
residual relaxation, a possibility of the alternative sets of
optical data for the complex index of refraction, an impact of
the surface roughness and patch potentials etc. (see below for the
references to each of these approaches).

The approaches of the second type admit that some serious
modifications of the Lifshitz theory might be necessary for a
resolution of the above problems. These approaches deal with
generalizations of the Lifshitz theory for the case of
configurations with nonplanar boundaries, with taken into account
spatial dispersion, screening effects etc. (the related references
are provided below). In this review, we consider both types of
attempts to find a resolution of the Casimir puzzle and the Casimir
conundrum including a very recent one which shows considerable
promise.

The review is organized as follows: In Section~2, we briefly
present the formulations of the Lifshitz theory in terms of real
and imaginary frequencies. In Section~3, a correlation is made
between the thermal Casimir forces for ideal and real metals.
The thermodynamic and experimental parts of the Casimir puzzle for
real metals are considered in Section~4. Section~5 is devoted to the
thermal Casimir force between ideal and real dielectrics and
Section~6 --- to the thermodynamic and experimental parts of the
Casimir conundrum. In Section~7, we summarize the main approaches
 to a resolution of the Casimir puzzle and the Casimir conundrum. In
Section~8, the new way toward resolving the Casimir puzzle is
considered. Section~9 is devoted to the discussion and in
Section~10 the reader will find our conclusions.

\section{The Lifshitz Theory of the Casimir Force}

There are different approaches to derivation of the Lifshitz theory starting
from the fluctuation-dissipation theorem of statistical physics \cite{5,6,47}
and quantum electrodynamics with continuity boundary conditions on the
boundary surfaces \cite{9,48,49}. For a discussion of the Casimir puzzle and
conundrum, we need only the expression for the Casimir free energy and force
in the configuration of two thick plates (semispaces) separated by a gap of width
$a$ and kept at temperature $T$ in thermal equilibrium with the environment.
In the framework of the Lifshitz theory it is assumed that the material of the
plates is described by the frequency-dependent dielectric permittivity
$\ve(\omega)$ and magnetic permeability $\mu(\omega)$. This means that the
original formulation of the Lifshitz theory does not take into account the
spatial dispersion. If the plate material is nonmagnetic, one should put
$\mu=1$.

\subsection{Formulation in Terms of Real Frequencies}

The immediate result of derivation using the fluctuation-dissipation theorem
is the following Lifshitz formula for the Casimir free energy per unit area
of the plates
\begin{eqnarray}
&&
{\cal F}(a,T)=\frac{\hbar}{4\pi^2}\int_0^{\infty}\!\kb d\kb
\int_0^{\infty}\!d\omega\coth\frac{\hbar\omega}{2k_BT}
\nonumber \\
&&~~~~~~~~~~~~~
\times
 {\rm Im}\left\{\ln\left[1-r_{\rm TM}^2(\omega,\kb)e^{-2aq}\right]+
\ln\left[1-r_{\rm TE}^2(\omega,\kb)e^{-2aq}\right]\right\}.
\label{eq3}
\end{eqnarray}
\noindent
Here, $\kb=\sqrt{k_1^2+k_2^2}$ is the projection of the wave vector on a plane
of the plates (it is perpendicular to the Casimir force), $\omega$ is the
frequency, $k_B$ is the Boltzmann constant and
$q=q(\omega,\kb)=(k_{\bot}^2-\omega^2/c^2)^{1/2}$.
The reflection coefficients for the transverse magnetic (TM) and transverse
electric (TE) polarizations of the electromagnetic field are given by
\begin{eqnarray}
&&
r_{\rm TM}(\omega,\kb)=
\frac{\ve(\omega)q(\omega,\kb)-k(\omega,\kb)}{\ve(\omega)q(\omega,\kb)+k(\omega,\kb)},
\nonumber \\
&&
r_{\rm TE}(\omega,\kb)=
\frac{\mu(\omega)q(\omega,\kb)-k(\omega,\kb)}{\mu(\omega)q(\omega,\kb)+k(\omega,\kb)},
\label{eq4}
\end{eqnarray}
\noindent
where
\begin{equation}
k(\omega,\kb)=\left[k_{\bot}^2-\ve(\omega)\mu(\omega)
\frac{\omega^2}{c^2}\right]^{1/2}.
\label{eq5}
\end{equation}
\noindent
It is seen that (\ref{eq4}) are the familiar Fresnel reflection coefficients of
classical electrodynamics.

The Casimir force per unit area of the plates is obtained from (\ref{eq3}) by
the negative differentiation with respect to $a$
\begin{eqnarray}
&&
{F}(a,T)=-\frac{\hbar}{2\pi^2}\int_0^{\infty}\!\kb d\kb
\int_0^{\infty}\!d\omega\coth\frac{\hbar\omega}{2k_BT}
\nonumber \\
&&~~~~~~~~~~~~~
\times
 {\rm Im}\left\{q\left[r_{\rm TM}^{-2}(\omega,\kb)e^{2aq}-1\right]^{-1}+
q\left[r_{\rm TE}^{-2}(\omega,\kb)e^{2aq}-1\right]^{-1}\right\}.
\label{eq6}
\end{eqnarray}

The integration in $\kb$ in (\ref{eq3}) and (\ref{eq6}) is performed from 0
to $\infty$ at any fixed $\omega$. In so doing the integration over
$\kb>\omega/c$ describes the contribution from the off-the-mass-shell
electromagnetic fluctuations which are also often called
{\it the evanescent waves}, as opposed to  {\it the propagating  waves}
which are characterized by $\kb\leqslant\omega/c$ and correspond to
the on-the-mass-shell fluctuations.

Note that the factor $\coth[\hbar\omega/(2k_BT)]$ in (\ref{eq3}) and
(\ref{eq6}) implies an origin of the Casimir force at $T\neq 0$ from
both the zero-point and thermal fluctuations of the electromagnetic
field. This factor arises from the free energy of an oscillator
\begin{equation}
\frac{\hbar\omega}{2}+k_BT\ln\left(1-e^{-\frac{\hbar\omega}{k_BT}}\right)=
k_BT\ln\left(2\sinh\frac{\hbar\omega}{2k_BT}\right)
\label{eq6a}
\end{equation}
\noindent
after an application of the Cauchy theorem \cite{9}.

For the off-the-mass-shell fluctuations the quantity $q$ is real.
Because of this, the contribution of these fluctuations to the Casimir free
energy and force can be calculated using (\ref{eq3}) and
(\ref{eq6}). However, for the on-the-mass-shell fluctuations the quantity $q$
becomes pure imaginary. As a result, both (\ref{eq3}) and
(\ref{eq6}) become the integrals of rapidly oscillating functions which plagues
their calculation. Because of this, another mathematically equivalent
formulation of the expressions (\ref{eq3}) and
(\ref{eq6}) was suggested \cite{5}.

\subsection{Formulation in Terms of Imaginary Matsubara Frequencies}

This formulation is based on the fact that  the dielectric permittivity is
a causal function which does not possess either poles or zeros in the upper
half plane of complex frequency. As a consequence, the dielectric permittivity
along the imaginary frequency axis takes the real values $\ve(\ix)$.
Finally, the equivalent formulation of (\ref{eq3}) is given by the Lifshitz
formula
\begin{eqnarray}
&&
{\cal F}(a,T)=\frac{k_BT}{2\pi}
\sum_{l=0}^{\infty}{\vphantom{sum}}^{\prime}\int_0^{\infty}\!\!\kb d\kb
\left\{\ln\left[1-r_{\rm TM}^2(\ix_l,\kb)e^{-2aq_l}\right]
\right.
\nonumber \\
&&~~~~~~~~~~~~~~~~~~~~~~~~~~+\left.
\ln\left[1-r_{\rm TE}^2(\ix_l,\kb)e^{-2aq_l}\right]\right\},
\label{eq7}
\end{eqnarray}
\noindent
where the prime on the summation sign divides the term with $l=0$ by 2 and
\begin{equation}
\xi_l=\frac{2\pi k_BTl}{\hbar}, \qquad
l=0,\,1,\,2,\,\ldots
\label{eq8}
\end{equation}
\noindent
are the Matsubara frequencies. The quantities $q_l$,  $r_{\rm TM}(\ix_l,\kb)$
and $r_{\rm TE}(\ix_l,\kb)$ are obtained from $q(\omega,\kb)$,
$r_{\rm TM}(\omega,\kb)$ and $r_{\rm TE}(\omega,\kb)$ defined above by a
replacement of $\omega$ with $\ix_l$.

In a similar way, the expression (\ref{eq6}) for the Casimir force takes the
form
\begin{eqnarray}
&&
{F}(a,T)=-\frac{k_BT}{\pi}
\sum_{l=0}^{\infty}{\vphantom{sum}}^{\prime}\int_0^{\infty}\!\!q_l\kb d\kb
\left\{\left[r_{\rm TM}^{-2}(\ix_l,\kb)e^{2aq_l}-1\right]^{-1}
\right.
\nonumber \\
&&~~~~~~~~~~~~~~~~~~~~~~~~~~+\left.
\left[r_{\rm TE}^{-2}(\ix_l,\kb)e^{2aq_l}-1\right]^{-1}\right\}.
\label{eq9}
\end{eqnarray}
\noindent
The expressions (\ref{eq7}) and (\ref{eq9}) are convenient for both analytical
and numerical calculations because $q_l$ is a real positive number at any $l$.

\section{Thermal Casimir Force between Ideal and Real Metals:
First Surprise}

The Casimir free energy (\ref{eq7}) and force (\ref{eq9}) are defined at nonzero
temperature $T$ and, therefore, are called {\it thermal}. In the limiting case
$T\to 0$ one has
\begin{equation}
k_BT\sum_{l=0}^{\infty}{\vphantom{sum}}^{\prime}\to
\frac{\hbar}{2\pi}\int_0^{\infty}\!d\xi,
\label{eq10}
\end{equation}
\noindent
the discrete Matsubara frequencies $\xi_l$ are replaced with the continuous
frequency $\xi$, $q_l$ is replaced with $q=(k_{\bot}^2+\xi^2/c^2)^{1/2}$
and we obtain from (\ref{eq7}) and (\ref{eq9}) the expressions for the Casimir
energy $E(a)$ and force $F(a)$ between real metal plates at zero temperature.

The ideal metal is perfect conductor at all frequencies including the zero
frequency, so that
\begin{equation}
r_{\rm TM}(\ix,\kb)=-r_{\rm TE}(\ix,\kb)=1.
\label{eq11}
\end{equation}
\noindent
This is in line with the prescription \cite{9a} which demands to consider
the limit of $\ve$ to infinity in (\ref{eq4}) first and the limit of frequency
to zero second.

Introducing into the Lifshitz formula (\ref{eq7}) written at $T=0$ with account
of (\ref{eq10}) the dimensionless variables $y=2aq$ and $\zeta=2a\xi/c$, and
taking into account (\ref{eq11}), one finds the Casimir energy between ideal
metal plates
\begin{equation}
E_{\rm IM}(a,0)=\frac{\hbar c}{16\pi^2a^3}\int_0^{\infty}\!d\zeta
\int_{\zeta}^{\infty}\!ydy\ln(1-e^{-y})=-\frac{\pi^2}{720}\,\frac{\hbar c}{a^3},
\label{eq12}
\end{equation}
\noindent
which is in agreement with the Casimir result (\ref{eq2}).

Similarly, the Lifshitz formula (\ref{eq9}) written at $T=0$ for ideal metal plates
results in
\begin{equation}
F_{\rm IM}(a,0)=-\frac{\hbar c}{16\pi^2a^4}\int_0^{\infty}\!d\zeta
\int_{\zeta}^{\infty}\!\frac{y^2dy}{e^{y}-1}=-\frac{\pi^2}{240}\,\frac{\hbar c}{a^4},
\label{eq13}
\end{equation}
\noindent
as was originally obtained by Casimir \cite{1} considering  the zero-point
oscillations of quantized electromagnetic field.

An application of the Lifshitz theory at nonzero temperature becomes more involved.
We illustrate it first in the case of large separation distances between the plates.
In this case all terms of (\ref{eq7}) and (\ref{eq9}) with $l\geqslant 1$ are
exponentially small and only the term with $l=0$ determines the total result.
The formal condition for the application of large separation limit is \cite{9}
\begin{equation}
a\gg\frac{\hbar c}{4\pi k_BT}.
\label{eq14}
\end{equation}
\noindent
At room temperature $T=300~$K this gives $a\gg0.64~\mu$m,  so that at
$a\geqslant 6~\mu$m one can safely use only the terms of (\ref{eq7}) and (\ref{eq9})
with $l=0$.

Then, for ideal metal plates, i.e., under a condition (\ref{eq11}), the Casimir free
energy and force at large separations are given by
\begin{eqnarray}
&&
{\cal F}_{\rm IM}(a,T)=\frac{k_BT}{2\pi}\int_{0}^{\infty}\!\kb d\kb
\ln(1-e^{-2a\kb})=-\frac{k_BT}{8\pi a^2}\zeta(3),
\nonumber \\
&&
{F}_{\rm IM}(a,T)=-\frac{k_BT}{\pi}\int_{0}^{\infty}\!
\frac{k_{\bot}^2 d\kb}{e^{2a\kb}-1}=-\frac{k_BT}{4\pi a^3}\zeta(3),
\label{eq15}
\end{eqnarray}
\noindent
where $\zeta(z)$ is the Riemann zeta function.

For real metals at large separations between the plates one might expect
a similar result. This expectation, however, lacks of support from calculations.
The point is that the low-frequency response of metals to the electromagnetic
field is described by the dissipative Drude model
\begin{equation}
\ve_D(\omega)=1-\frac{\omega_p^2}{\omega[\omega+{\rm i}\gamma(T)]}, \qquad
\ve_D(\ix)=1+\frac{\omega_p^2}{\xi[\xi+\gamma(T)]},
\label{eq16}
\end{equation}
\noindent
where $\omega_p$ is the plasma frequency and $\gamma(T)\ll\omega_p$ is the
relaxation parameter.

For the nonmagnetic metals one has from (\ref{eq5})
\begin{equation}
\lim_{\xi\to 0}\xi^2\ve_D(\ix)=0,\qquad
\lim_{\xi\to 0}k(\ix,\kb)=\kb.
\label{eq17}
\end{equation}
\noindent
Then, from (\ref{eq4}) we obtain
\begin{equation}
r_{{\rm TM},D}(0,\kb)=1, \qquad
r_{{\rm TE},D}(0,\kb)=0,
\label{eq18}
\end{equation}
\noindent
which is in contradiction with (\ref{eq11}) valid for an ideal metal.
As a result, in the limiting case of large separations the terms of
(\ref{eq7}) and (\ref{eq9}) with $l=0$ are equal to
\begin{eqnarray}
&&
{\cal F}_{D}(a,T)=\frac{k_BT}{4\pi}\int_{0}^{\infty}\!\kb d\kb
\ln(1-e^{-2a\kb})=-\frac{k_BT}{16\pi a^2}\zeta(3),
\nonumber \\
&&
{F}_{D}(a,T)=-\frac{k_BT}{2\pi}\int_{0}^{\infty}\!
\frac{k_{\bot}^2 d\kb}{e^{2a\kb}-1}=-\frac{k_BT}{8\pi a^3}\zeta(3),
\label{eq19}
\end{eqnarray}
\noindent
i.e., by the factor of 2 smaller in magnitude than for an ideal metal
in (\ref{eq15}).

It is interesting that if, instead of (\ref{eq16}), the dissipationless
plasma model
\begin{equation}
\ve_p(\omega)=1-\frac{\omega_p^2}{\omega^2}, \qquad
\ve_D(\ix)=1+\frac{\omega_p^2}{\xi^2},
\label{eq20}
\end{equation}
\noindent
is used, one finds
\begin{equation}
\lim_{\xi\to 0}\xi^2\ve_p(\ix)=\omega_p^2 ,\qquad
\lim_{\xi\to 0}k(\ix,\kb)=\left(k_{\bot}^2+\frac{\omega_p^2}{c^2}\right)^{1/2}.
\label{eq21}
\end{equation}
\noindent
This leads to the following values of the reflection coefficients at zero frequency:
\begin{equation}
r_{{\rm TM},p}(0,\kb)=1, \qquad
r_{{\rm TE},p}(0,\kb)=
\frac{c\kb-\sqrt{c^2k_{\bot}^2+\omega_p^2}}{c\kb+\sqrt{c^2k_{\bot}^2+\omega_p^2}}.
\label{eq22}
\end{equation}

For the Casimir free energy and force defined using the plasma model at large
separations, the Lifshitz theory leads to
\begin{eqnarray}
&&
{\cal F}_{D}(a,T)=-\frac{k_BT}{16\pi a^2}\zeta(3)+
\frac{k_BT}{4\pi}\int_{0}^{\infty}\!\kb d\kb
\ln[1-r_{{\rm TE},p}^2(0,\kb)e^{-2a\kb}],
\nonumber \\
&&
{F}_{D}(a,T)=-\frac{k_BT}{8\pi a^3}\zeta(3)-
\frac{k_BT}{2\pi}\int_{0}^{\infty}\!
\frac{k_{\bot}^2 d\kb}{r_{{\rm TE},p}^{-2}(0,\kb)e^{2a\kb}-1},
\label{eq23}
\end{eqnarray}
\noindent
where $r_{{\rm TE},p}(0,\kb)$ is defined in (\ref{eq22}). Taking into account that
in the limiting case of ideal metal ($\omega_p\to\infty$) it holds
\begin{equation}
\lim_{\omega_p\to\infty}r_{{\rm TE},p}(0,\kb)=-1
\label{eq24}
\end{equation}
\noindent
in accordance to (\ref{eq22}), we conclude that the results (\ref{eq23})
 obtained for real metal described by the plasma model join smoothly with the
respective results (\ref{eq15}) obtained for ideal metal plates.
This could be considered as an advantage of the plasma model, as compared to
the Drude one.

It should be remembered, however, that in the region of
quasistatic frequencies the relaxation properties of conduction electrons
in metals disregarded by the plasma model play an important role and should
be taken into account. Moreover, according to the Bohr-van~Leeuwen theorem
of classical statistical physics the TE electromagnetic fields should decouple
from matter in the classical limit. According to the results of \cite{49a},
this leads to the zero value of $r_{\rm TE}$ in this limit (see, however,
Section~7 concerning the role of quantum fluctuations off the mass shell at
all separations).

The complexity to this problem was added by a calculation of the thermal
correction to the Casimir force between metallic plates described by the Drude
model \cite{10}. It turned out that this correction takes relatively large in
magnitude negative values and, thus, decreases the magnitude of the Casimir
force which is a counterintuitive result. We recall that for the ideal metal
plates at room temperature the thermal correction to the zero-temperature
force at separations below $1~\mu$m is negligibly small. This result is
justified by a smallness of parameter $T/T_{\rm eff}$ where
$k_BT_{\rm eff}=\hbar c/(2a)$.

\begin{figure}[!b]
\centering
\vspace*{-11.cm}
\hspace*{-1.6cm}\includegraphics[width=16 cm]{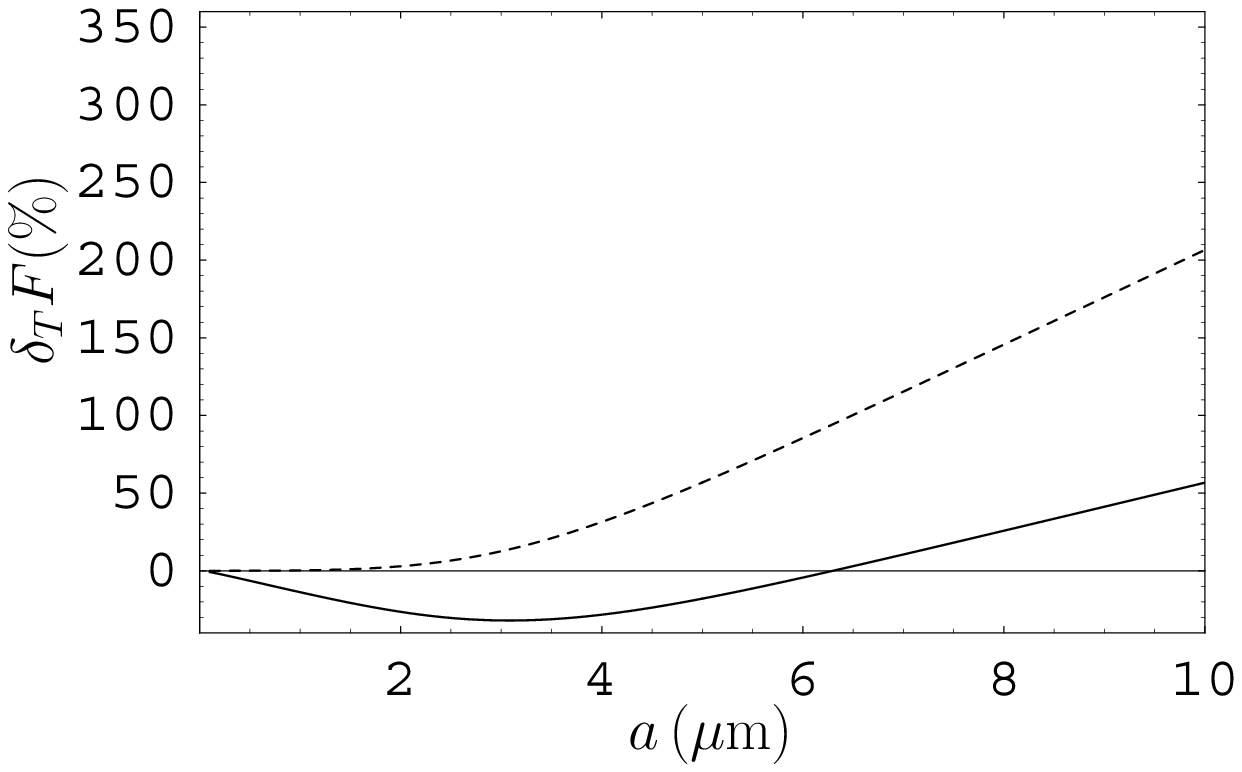}
\vspace*{-5.cm}
\caption{The relative thermal corrections for the Casimir force
between gold plates computed in the framework of the Lifshitz theory
at $T=300~$K using the Drude  and the plasma models  are shown as the
functions of separation by the solid and dashed lines, respectively.
\label{fg1}}
\end{figure}

To illustrate a fundamental difference between the thermal effects in the
Casimir force predicted by the Drude and plasma models  we consider the quantity
\begin{equation}
\delta_T F_{D(p)}(a,T)=\frac{F_{D(p)}(a,T)-F_{D(p)}(a,0)}{F_{D(p)}(a,T)}
\label{eq25}
\end{equation}
\noindent
computed by (\ref{eq9}) using the Drude (\ref{eq16}) and the plasma (\ref{eq20})
dielectric permittivities of gold with $\omega_p=9.0~$eV and $\gamma(T)=0.035~$eV
at $T=300~$K. The computational results in percent
are shown in Figure~\ref{fg1}
as the functions of separation between the plates by the solid and dashed lines
for the Drude and plasma models, respectively.

As is seen in Figure~\ref{fg1}, the Lifshitz theory using the Drude model predicts
rather large in magnitude thermal correction to the Casimir force of the opposite
sign to the zero-temperature contribution up to the separation of $6.3~\mu$m
where it vanishes \cite{10}. For example, at $a=500~$nm, 700~nm, and $1~\mu$m
this correction is equal to --6.4\%, --9.4\%, and --13.8\%, respectively
(see the solid line). In contrast to this, the Lifshitz theory using the plasma
model predicts the thermal correction of the same sign as the zero-temperature
Casimir force which magnitude is in qualitative agreement with that for ideal
metal plates, i.e., very small at separations below $1~\mu$m \cite{11}.
Thus, at separations $a=500~$nm and $1~\mu$m we have $\Delta_T F_p=0.058$\% and
0.29\%, respectively. These results remain almost unchanged when one uses not
the simple Drude and plasma models (\ref{eq16}) and (\ref{eq20}) but the
dielectric permittivities obtained from the measured optical data for the
complex index of refraction of gold \cite{50}, extrapolated by the
Drude and plasma models down to zero frequency where the optical data are
unavailable.

Starting from 2000, the problems in the Lifshitz theory have attracted considerable
attention. The surprising thing is that the use of the well-tested Drude model
leads to somewhat strange and unexpected results whereas the results obtained
using the seemingly inapplicable plasma model are quite reasonable from the
theoretical point of view. As is seen in the next section, at a later time this
problem was further aggravated both theoretically and experimentally.

\section{Thermodynamic and Experimental Parts of the Casimir Puzzle for
Real Metals}

As was mentioned in Section~1, first measurements of the Casimir force were not
enough precise and demonstrated only a qualitative agreement with theoretical
predictions of the Lifshitz theory. With an advent of micromechanical devices,
it has become possible to perform precise and reproducible measurements of small
forces at separations below a micrometer. At the same time, quick progress in
analytical calculations using the powerful computer means made more accessible
an investigation of the asymptotic behaviors of complicated mathematical
expressions. Taken together, these achievements gave the possibility to
compare in more detail the  predictions of the Lifshitz theory with the
measurement data and correlate  its formalism  with the fundamental laws of
thermodynamics. Below we consider both the experimental and thermodynamic
aspects of the Lifshitz theory. It is shown that they constitute two parts of the
grave problem which received the name of {\it the Casimir puzzle}.
We begin with the thermodynamic part.

\subsection{The Casimir Entropy for Metallic Plates and the Nernst Heat
Theorem}

The Casimir free energy per unit area of two parallel metallic plates kept at
temperature $T$ in thermal equilibrium with the environment is expressed by the
Lifshitz formula (\ref{eq7}). Taking into consideration that there are problems
in the Lifshitz theory considered in Section~3, it is interesting to perform
some kind of thermodynamic checking for a behavior of the Casimir free energy and
respective entropy at vanishing temperature when the Drude and the plasma models
of the dielectric response are used. It is well known that in accordance to the
third law of thermodynamics (the Nernst heat theorem) with vanishing temperature
the entropy of a closed system in the state of thermal equilibrium must go to
zero  or to some universal constant which does not depend on the parameters of
this system \cite{51,52}.

To check whether or not the Casimir entropy satisfies the Nernst heat theorem
the metal of the plates with perfect crystal lattice was considered \cite{12,13,14}.
Perfect crystal lattice is a basic theoretical model of condensed matter physics
possessing the nondegenerate ground state. For perfect crystal lattices the
relaxation parameter of the Drude model  (\ref{eq16}) goes to zero with vanishing
temperature as $\gamma(T)\sim T^2$. This dependence is caused by the electron-electron
scattering and is followed at all $T$ below the temperature of liquid helium \cite{53}.
In fact at higher $T$ the relaxation parameter decreases following the power law with
different powers but the condition $\gamma(T)\ll\xi_1(T)$ is satisfies in all cases.
This is shown in \cite{14}.

For obtaining the asymptotic behavior of the Casimir free energy ${\cal F}_D$ when the
Drude model is used one should substitute $\ve(\ix)=\ve_D(\ix)$ in (\ref{eq7}).
It is convenient to separate the term with $l=0$ and expand the sum of all other terms
in powers of the small parameter  $\gamma(T)/\xi_1(T)\ll 1$. Then, calculating the
Casimir entropy as
\begin{equation}
S_{D(p)}(a,T)=-\frac{\partial {\cal F}_{D(p)}(a,T)}{\partial T},
\label{eq26}
\end{equation}
\noindent
one can  present $S_D$ in the form \cite{9,14}
\begin{equation}
S_{D}(a,T)=S_p(a,T)+
\frac{k_B}{4\pi}\int_0^{\infty}\!\!\kb d\kb\ln\left[1-r_{{\rm TE},p}^2(0,\kb)
e^{-2a\kb}\right] +O\left(T\ln\frac{T}{T_{\rm eff}}\right).
\label{eq27}
\end{equation}

The Casimir entropy $S_p$ calculated using the plasma model was found \cite{13,14}
as a perturbation expansion in powers of the small parameter $\delta_0/a\ll 1$
where $\delta_0=c/\omega_p$ is the skin depth of a metal
\begin{eqnarray}
&&
S_p(a,T)=\frac{k_B^3T^2}{\pi\hbar^2c^2}\left\{
\vphantom{\left(\frac{\delta_0}{a}\right)^{\!\!2}\,\frac{20\zeta(5)a^2k_B^2T^2}{\hbar^2c^2}}
\frac{3}{2}\zeta(3)-
\frac{4\pi^3ak_BT}{45\hbar c}+
\frac{\delta_0}{a}\left[3\zeta(3)-\frac{16\pi^3ak_BT}{45\hbar c}\right]\right.
\nonumber\\
&&~~~~~~~~~~~~~~~~~~~~~~
\left.
-\left(\frac{\delta_0}{a}\right)^{\!\!2}\,\frac{20\zeta(5)a^2k_B^2T^2}{\hbar^2c^2}\right\}.
\label{eq28}
\end{eqnarray}
\noindent
Note that the contribution of the zeroth order in $\delta_0/a$ in (\ref{eq28})
coincides with the Casimir entropy for the ideal metal plates at low
temperature \cite{54}. It is seen that
\begin{equation}
\lim_{T\to 0}S_p(a,T)=S_p(a,0)=0,
\label{eq29}
\end{equation}
\noindent
i.e., the Lifshitz theory using the plasma model satisfies the Nernst heat theorem.

This is not the case, however, when the Drude model is used. From (\ref{eq27}) with
the help of (\ref{eq29}) one obtains
\begin{equation}
\lim_{T\to 0}S_D(a,T)=S_D(a,0)=\frac{k_B}{4\pi}\int_0^{\infty}\!\!\kb d\kb
\ln\left[1-r_{{\rm TE},p}^2(0,\kb) e^{-2a\kb}\right].
\label{eq30}
\end{equation}
\noindent
Expanding this expression in powers of the small parameter $\delta_0/a$, we
arrive at
\begin{equation}
S_D(a,0)=-\frac{k_B\zeta(3)}{16\pi a^2}\left[1-4\frac{\delta_0}{a} +
12\left(\frac{\delta_0}{a}\right)^2-\,\ldots\right]<0.
\label{eq31}
\end{equation}
\noindent
The quantities (\ref{eq30}) and (\ref{eq31}) depend on the parameters of a system
such as the separation between the plates and the plasma frequency which means that
the Nernst heat theorem is violated.
Note that in \cite{54a,54b} this violation was attributed to the role of eddy currents
which explain big difference between theoretical predictions for the Casimir force
when the Drude and the plasma models are used. According to \cite{54a,54b}, the nonzero
entropy \\(\ref{eq31}) at $T=0$ is explained by an initiation of the correlated glassy
(i.e., nonequilibrium) state.

Here we considered the parallel plates made of a nonmagnetic metal. The same results are,
however, valid for other geometric configurations and for magnetic metals
\cite{15,16,17,18}. These results are really puzzling because the Lifshitz theory
arrives to a conflict with the fundamental law of thermodynamics when using the
Drude response function in the frequency region where it is well applicable.
It is no less surprising that an agreement of the Lifshitz theory with
thermodynamics is restored when the plasma response function  is used in the
region of quasistatic frequencies, i.e., outside of its application region.


\subsection{The Lifshitz theory Confronts Experiments with Metallic Test Bodies}

First precise measurements of the Casimir force allowing a discrimination between
theoretical predictions using the plasma and Drude response functions have been
performed by means of micromechanical torsional oscillator in the configuration
of an Au-coated sapphire sphere above an Au-coated polysilicon plate
\cite{19,20,21,22}. The thicknesses of Au coatings were sufficiently large in
order the test bodies could be considered as all-gold. The dielectric permittivity
of Au along the imaginary frequency axis was found by means of the Kramers-Kronig
relations using the available tabulated optical data of Au \cite{50} extrapolated
down to zero frequency either by the Drude or the plasma model \cite{55}.

As discussed in Section~2, the Lifshitz theory was formulated for a configuration
of two parallel plates. It turned out, however, that the sphere-plate configuration
is much more convenient for precise measurements of the Casimir interaction.
Taking into account that in the early 21st century the exact expression for the
Casimir force between a sphere and a plate was not available, the comparison
between experiment and theory have been made using the so-called
{\it proximity force approximation} (PFA) \cite{9,55,56}.

According to PFA, the Casimir force between a sphere and a plate $F_{sp}$ is
approximately equal to
\begin{equation}
F_{sp}(a,T)=2\pi R{\cal F}(a,T),
\label{eq32}
\end{equation}
\noindent
where ${\cal F}(a,T)$ is the Casimir free energy per unit area of two parallel
plates defined in \\(\ref{eq7}), $a$ is the minimum sphere-plate separation, and
$R$ in the sphere radius. The error of this approximation was not known
precisely at that time but it was expected to be of the order of $a/R$.
Taking into account that the measurements \cite{19,20,21,22} were performed at $a<1~\mu$m
with the sphere radius $R\approx 150~\mu$m, this error does not exceed 0.7\%.
The most precise experiments on measuring the Casimir interaction were performed
in the dynamic mode where not the Casimir force but its gradient
$F_{sp}^{\prime}$ has been measured. By differentiating (\ref{eq32}) with respect
to $a$, the force gradient is expressed by the Lifshitz formula (\ref{eq9}) for
the force per unit area of two parallel plates
\begin{equation}
F_{sp}^{\prime}(a,T)\equiv\frac{\partial F_{sp}(a,T)}{\partial a} =-2\pi R{F}(a,T).
\label{eq33}
\end{equation}

Another point which was taken into account when comparing experiment with theory
is an impact of the surface roughness. The roughness profiles on both interacting surfaces
have been examined using an atomic force microscope. The heights of the maximum
roughness peaks were found to be much less than the minimum separation between
a sphere and a plate. In this case the Casimir force $F(a,T)$ between  two
parallel plates with account of surface roughness was found using an additive
method of geometrical averaging of the forces (\ref{eq9}) calculated at the local
separations $a_i$ over the profiles of both surfaces \cite{9,55,57}.
The gradient of the Casimir force with account of surface roughness was then
calculated using (\ref{eq33}).

\begin{figure}[!b]
\centering
\vspace*{-2.2cm}
\hspace*{-1.6cm}\includegraphics[width=16 cm]{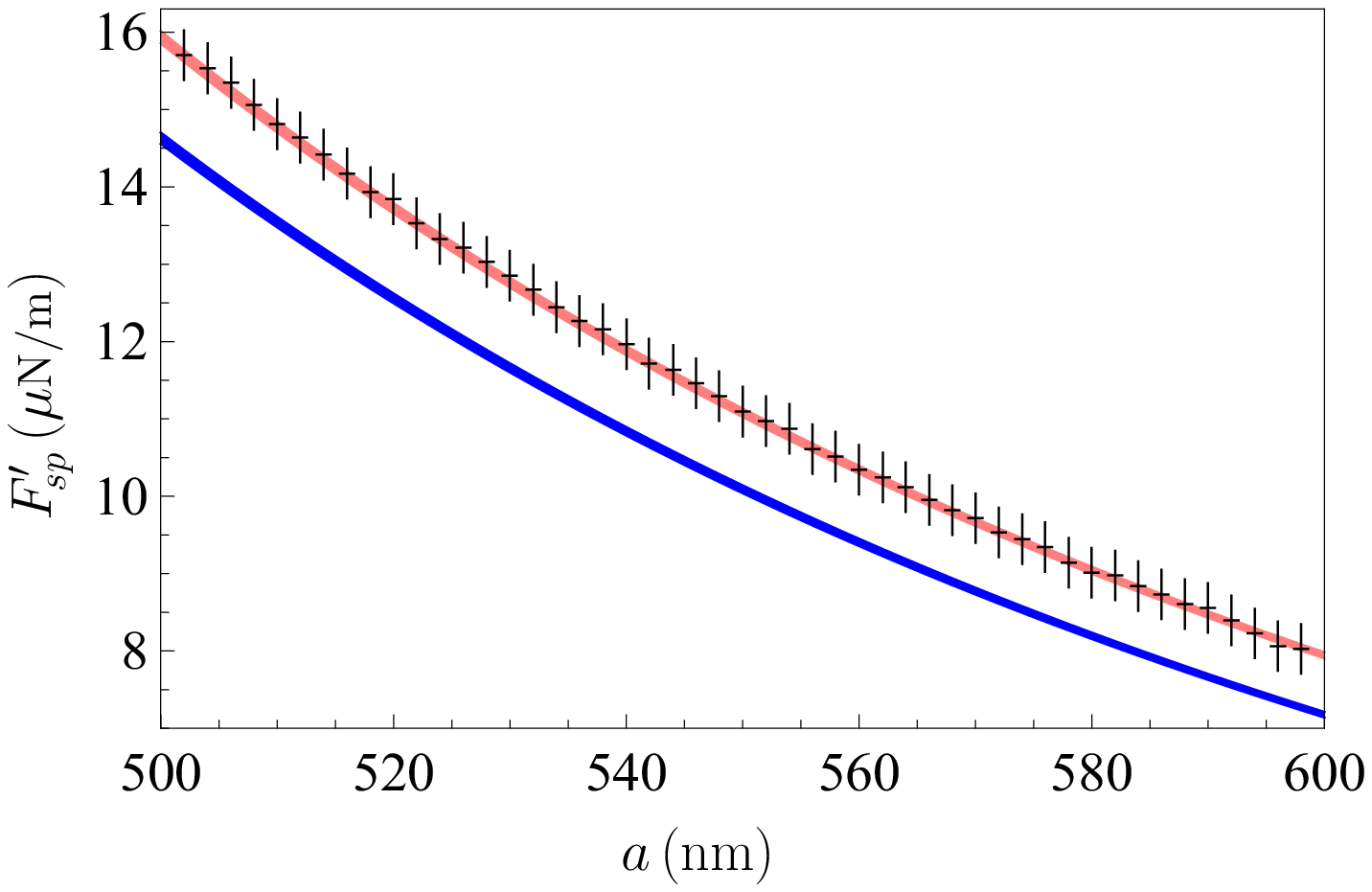}
\vspace*{-13.cm}
\caption{The gradients of the Casimir force between an Au-coated sphere and
an Au-coated plate measured by means of micromechanical torsional oscillator
\cite{21} (crosses) are compared with theoretical predictions of the Lifshitz
theory using the Drude and plasma extrapolations of the optical data of Au
(blue and red bands, respectively).
\label{fg2}}
\end{figure}

In Figure~\ref{fg2}, we present the typical comparison between the measurement data
obtained using a micromechanical torsional oscillator in high vacuum \cite{21}
and theoretical predictions of the Lifshitz theory. The predictions found with
the Drude and plasma extrapolations of the optical data are shown by the blue and
red bands, respectively, as the functions of separation. The widths of the bands
reflect the total theoretical error  which includes inaccuracies of the PFA,
errors in the optical data and in the sphere radius. The mean experimental data
are shown as crosses. The arms of the crosses undicate the total experimental errors.
All errors are found at the 95\% confidence level. Note that in \cite{21} the
comparison between experiment and theory was made in terms of the effective Casimir
pressure between two parallel plates which is equivalently presented here in terms
of the originally measured gradients of the Casimir force between a sphere and
a plate.

As is seen in Figure~\ref{fg2}, the Lifshitz theory using the Drude model for an
extrapolation of the optical data is experimentally excluded. The same result
was obtained over the separation region from 160 to 750~nm. From Figure~\ref{fg2}
it is apparent that the Lifshitz theory using the plasma model for an extrapolation
is in a very good agreement with the measurement data. This is the case also over
the entire measurement range \cite{21}.

As the second example of comparison between experiment and theory, we consider
experiment on measuring the gradient of the Casimir force between a sphere of
$R\approx 62~\mu$m radius and a plate with coated by a magnetic metal Ni surfaces
\cite{24,25}. These experiments were performed in high vacuum using an atomic
force microscope operated in the dynamic mode. The metallic coatings of the test
bodies were not magnetized, so that the magnetic force did not contribute to the
measurement results. It should be recalled also that at room temperature the
magnetic permeability $\mu(\ix)$ falls to unity at frequencies much below the
first Matsubara frequency \cite{58}. Because of this, the magnetic properties of
Ni make an impact on the Casimir force only through the term of (\ref{eq9}) with
$l=0$ where $\mu(0)=110$ whereas in all terms with $l\geqslant 1$ one should
put $\mu(\ix_l)=1$.

\begin{figure}[!b]
\centering
\vspace*{-2.2cm}
\hspace*{-1.6cm}\includegraphics[width=16 cm]{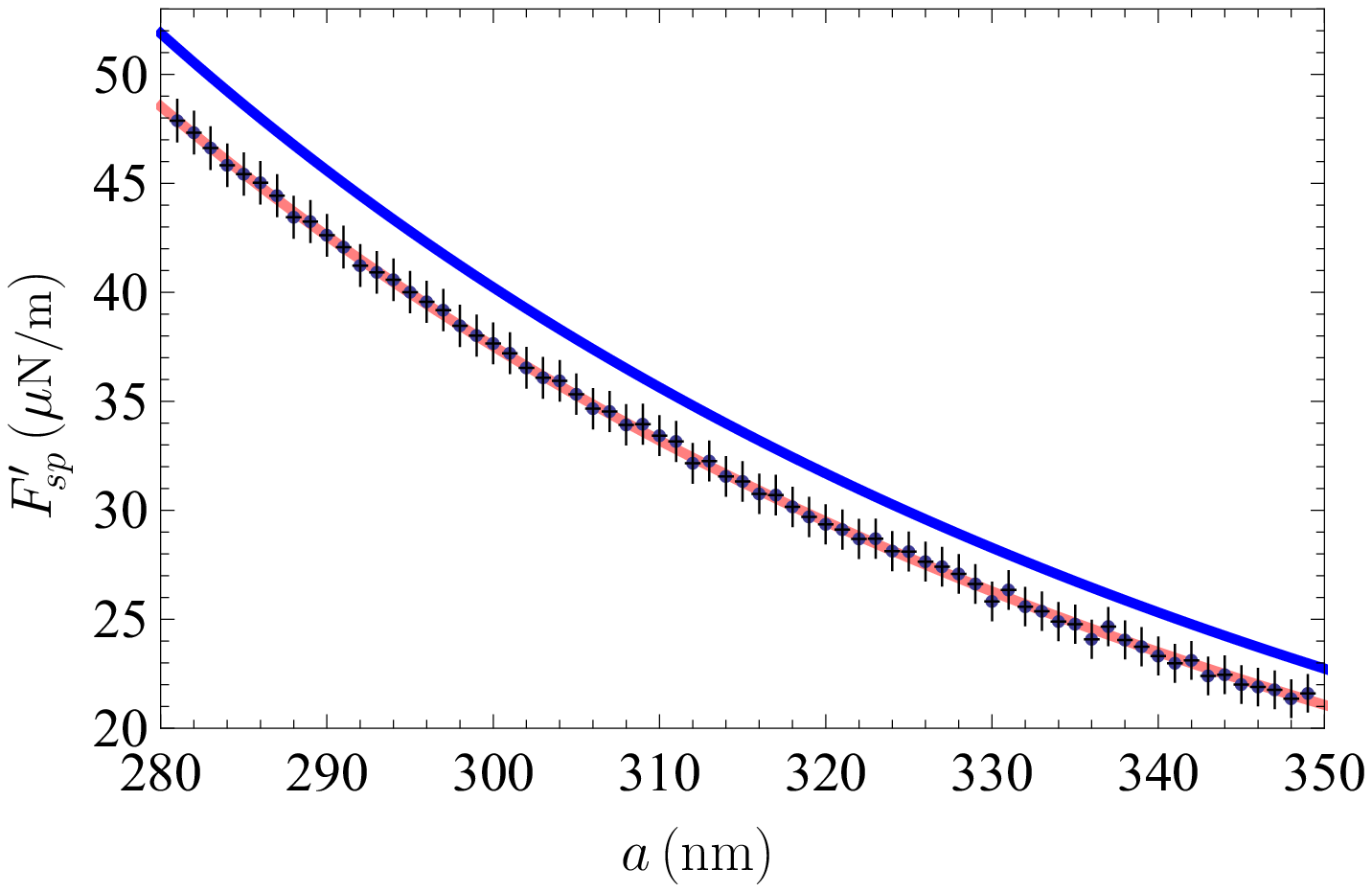}
\vspace*{-13.cm}
\caption{The gradients of the Casimir force between a Ni-coated sphere and
a Ni-coated plate measured by means of dynamic atomic force microscope
\cite{24} (crosses) are compared with theoretical predictions of the Lifshitz
theory using the Drude and plasma extrapolations of the optical data of Ni
(blue and red bands, respectively).
\label{fg3}}
\end{figure}
The comparison between an experimen and theory was made in the same way as above using
the optical data of Ni \cite{50} extrapolated to zero frequency either by the Drude or
by the plasma model, the proximity force approximation, and the additive approach to an
account of the surface roughness. In Figure~\ref{fg3}, the predictions of the Lifshitz
theory using the Drude and plasma models are shown as the blue and red bands,
respectively, whereas the measurement data are shown as crosses with all errors
determined at the 67\% confidence level \cite{24}. It is seen that the Lifshitz theory
using the Drude model is excluded by the measurement data which are in good agreement
with the same theory using the plasma model. Altogether the Lifshitz theory using the
Drude model was
excluded over the separation region from 223~nm (the minimum separation of this
experiment) to 420~nm.

It should be noted that measurements of the Casimir interaction between magnetic
metals have an important difference as compared to experiments using the gold test
bodies. As is seen in Figure~\ref{fg2}, the predictions of the Lifshitz theory
using the Drude model (the blue band) are situated below the experimental crosses.
This means that in order to bring these predictions in agreement with the data some
additional {\it attractive} force would be needed. By contrast, for the magnetic
test bodies the predictions of the Lifshitz theory
using the Drude model lie above the measurement data. An agreement between them
could be reached only at the expense of some extra {\it repulsive} force.
Thus, both sets of experiments make any attempt to bring the Lifshitz theory
combined with the Drude model in agreement with measurements a challenging
task (see Section~7).

{}From Figures \ref{fg2} and \ref{fg3} it is seen that, although the Lifshitz theory
using the Drude model is clearly excluded by the measurement data, its predictions
differ from these data (and from the predictions of the same theory using the
plasma model) by only a few percent of the force gradient. It was shown, however,
that using the differential force measurements, originally suggested for searching
the Yukawa-type corrections to Newton's gravitational law \cite{59,60}, it is
possible to increase a difference between the predictions of the Lifshitz theory
using the Drude and plasma response functions by the several orders of magnitude
\cite{61,62}. Using this idea, the differential Casimir force was measured
between a Ni-coated sphere and Au and Ni sectors covered with an Au overlayer
\cite{26}. A presence of the overlayer makes almost equal the contributions of all
nonzero Matsubara frequencies in the positions of a sphere above the Au and Ni sectors.
As a result, the differences in the Casimir forces acting on a sphere in these positions,
calculated by the Lifshitz theory using the Drude and the plasma model extrapolations
of the optical data, diverge by up to a factor of 1000. The measurement results
for the differential Casimir force were found in good agreement with the predictions
of the Lifshitz theory using the plasma model extrapolations
of the optical data. The alternative predictions using the Drude model, which differ
from the confirmed ones by up to a factor of 1000, were found to be out of all
proportion to the actual size of the measured differential force.

The comparison between the predictions of the Lifshitz theory and the measurement
data of precise experiments presented above is puzzling because the commonly used
and considered as correct Drude response function failed, whereas the evidently
inapplicable plasma response function demonstrated that it can be used successfully.

\section{The Casimir Force between Ideal and Real Dielectrics}

The ideal dielectric or insulator is a material which does not conduct an electric
current at any temperature. The dielectric permittivity of an ideal dielectric
at the pure imaginary Matsubara frequencies can be presented in the oscillator
form \cite{63}
\begin{equation}
\ve(\ix_l)=1+\sum_{j=1}^{K}\frac{g_j}{\omega_j^2+\xi_l^2+\gamma_j\xi_l},
\label{eq34}
\end{equation}
\noindent
where $\omega_j$ are the oscillator frequencies (all of them are nonzero),
$\gamma_j$ are the relaxation parameters, $g_j$ are oscillators strengths,
and $K$ is the number of oscillators.

At zero Matsubara frequency, $\xi_0=0$, the permittivity of an ideal dielectric
takes the finite value
\begin{equation}
\ve(0)=1+\sum_{j=1}^{K}\frac{g_j}{\omega_j^2}.
\label{eq35}
\end{equation}
\noindent
By putting $\omega=\ix=0$ in (\ref{eq4}) for an ideal dielectric with no magnetic
properties ($\mu=1$) one obtains
\begin{equation}
r_{\rm TM}(0,\kb)=\frac{\ve(0)-1}{\ve(0)+1},\qquad
r_{\rm TE}(0,\kb)=0.
\label{eq36}
\end{equation}
\noindent
As is seen in (\ref{eq36}), the value of the TE reflection coefficient is the same
as for the Drude model in (\ref{eq18}) whereas the value of the TM one is different.

According to Section~3, the zero-frequency terms of the Lifshitz formulas determine
the behaviors of the Casimir free energy and force at large separations.
Substituting (\ref{eq36}) in (\ref{eq7}) and integrating, we find the Casimir free
energy per unit area of the plates made of an ideal dielectric
\begin{equation}
{\cal F}_{\rm ID}(a,T)=\frac{k_BT}{4\pi}\int_{0}^{\infty}\!\!\kb d\kb
\ln\left[1-r_{\rm TM}^2(0,\kb)e^{-2a\kb}\right]=
-\frac{k_BT}{16\pi a^2}Li_3[r_{\rm TM}^2(0)],
\label{eq37}
\end{equation}
\noindent
where $r_{\rm TM}(0,\kb)\equiv r_{\rm TM}(0)$ is defined in (\ref{eq36}) and
$Li_3(z)$ is the polylogarithm function.

In a similar way, substituting (\ref{eq36}) in (\ref{eq9}), one finds the
asymptotic behavior of the Casimir force per unit area
between two ideal dielectric plates
at large separations
\begin{equation}
{F}_{\rm ID}(a,T)=-\frac{k_BT}{2\pi}\int_{0}^{\infty}
\frac{k_{\bot}^2 d\kb}{r_{\rm TM}^{-2}(0,\kb)e^{2a\kb}-1}=
-\frac{k_BT}{8\pi a^3}Li_3[r_{\rm TM}^2(0)].
\label{eq38}
\end{equation}

Real dielectrics are also characterized by the zero electric conductivity at zero
temperature. This property differentiates them from metals which have a nonzero
conductivity at any temperature. In this regard all materials are either dielectrics
or metals \cite{64}. There are several types of real dielectrics but all of them
have some nonzero conductivity at $T>0$.

The dielectric permittivity of a real dielectric at temperature $T$ considered at
the pure imaginary Matsubara frequencies can be presented in the form \cite{50}
\begin{equation}
\ve_{\rm RD}(\ix_l)=\ve(\ix_l)+\frac{4\pi\sigma_0(T)}{\xi_l},
\label{eq39}
\end{equation}
\noindent
where $\ve(\ix_l)$ is determined in (\ref{eq34}) and $\sigma_0$ is the
temperature-dependent static conductivity. This  static conductivity of dielectric
material vanishes with temperature exponentially fast
\begin{equation}
\sigma_0(T)\sim\exp\left(-\frac{\Delta}{2k_BT}\right),
\label{eq40}
\end{equation}
\noindent
where $\Delta$ is the band gap. It is convenient to call the dielectric material
{\it an insulator} if $\Delta > 2-3~$eV and
{\it an intrinsic semiconductor} if $\Delta < 2-3~$eV \cite{64}.

Substituting (\ref{eq39}) in (\ref{eq4}) one obtains for real dielectrics the
following expressions for the reflection coefficients at zero frequency:
\begin{equation}
r_{\rm TM}(0,\kb)=1,\qquad
r_{\rm TM}(0,\kb)=0,
\label{eq41}
\end{equation}
\noindent
which coincide with (\ref{eq18}) obtained for metals described by the Drude
model but are different from (\ref{eq36}) obtained for ideal dielectrics.
Thus, the Casimir free energy and force for dielectric
plates at sufficiently
large separations are given by expressions (\ref{eq19}) obtained in Section~3
for metallic plates described by the Drude model.

In view of this results one can suggest that an inclusion of the dc conductivity in
calculation of the Casimir free energy and force between dielectric plates plays the
same role as an inclusion of the relaxation properties of conduction electrons for
the metallic ones. This guess is confirmed by calculations of the relative thermal
correction to the Casimir force between dielectric plates defined as
\begin{equation}
\delta_T F_{\rm RD(ID)}(a,T)=\frac{F_{\rm RD(ID)}(a,T)-
F_{\rm RD(ID)}(a,0)}{F_{\rm RD(ID)}(a,0)}.
\label{eq42}
\end{equation}
\noindent
Computations have been performed for a fused silica using the optical
data for its complex index of refraction \cite{50} with $\ve(0)=3.81$ (ideal
dielectric) and with taken into account typical value of the dc conductivity of
fused silica at room temperature $\sigma_0=29.7~\mbox{s}^{-1}$ (real
dielectric). Note that the computational results for real fused silica do not depend
on the value of $\sigma_0$ but only on the fact that it is not equal to zero.

\begin{figure}[!h]
\centering
\vspace*{-11.cm}
\hspace*{-1.6cm}\includegraphics[width=16 cm]{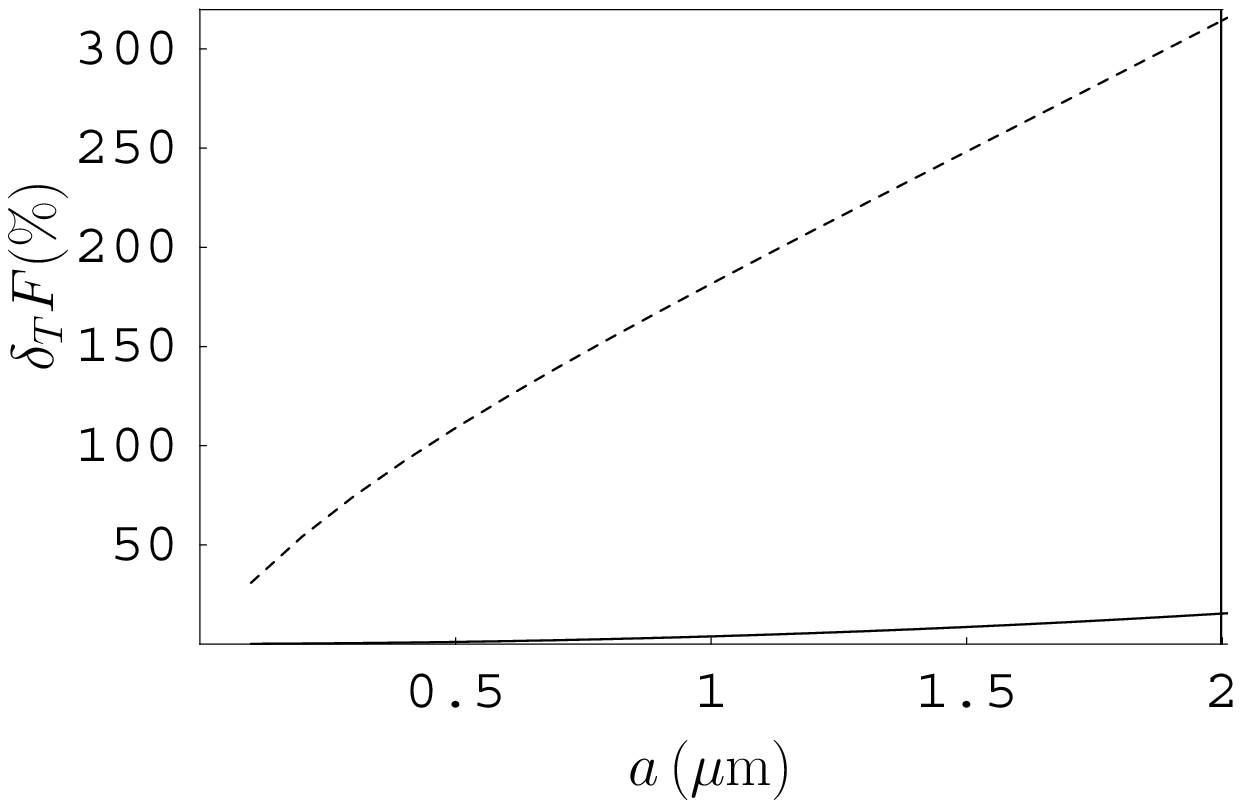}
\vspace*{-5.cm}
\caption{The relative thermal corrections for the Casimir force
between fused silica plates computed in the framework of the
Lifshitz theory at $T=300~$K  with taken into account and omitted dc
conductivity are shown as the
functions of separation by the dashed and solid lines, respectively.
\label{fg4}}
\end{figure}

In Figure~\ref{fg4}, the predictions of the Lifshitz theory for the thermal
correction (\ref{eq42}) in percent for the real and ideal fused silica (i.e., with
included and omitted dc conductivity) at $T=300~$K  are shown as the functions
of separation between the plates by the dashed and solid lines, respectively.
As is seen in this figure, at short separations the thermal correction is
reasonably small if the dc conductivity is omitted in computations (see the
solid line). Thus, at $a=1$ and $2~\mu$m it is equal to only 3.9\% and 15.4\%,
respectively. The situation reverses if the dc conductivity is taken into
account in computations (see the dashed line). In this case the relative
thermal correction (\ref{eq42}) is equal to 182\% and 314\% at the same
respective separations.

It should be noted that so big difference between the predictions of the
Lifshitz theory with omitted and included dc conductivity for dielectrics
arises from a difference between the TM reflection coefficients in (\ref{eq36})
and (\ref{eq41}). We recall that a similar difference between predictions of the
same theory for metallic plates described by the Drude and plasma models is due
to a difference between the TE reflection coefficients in (\ref{eq18}) and (\ref{eq22}).
In any case, the above results suggest that the grave problems met by the Lifshitz
theory in the case of metals (see Section~4) may appear for dielectrics as well.
This conclusion is confirmed in the next section.

\section{Thermodynamic and Experimental Parts of the  Casimir Conundrum
for Dielectrics}

Consideration of the thermal Casimir force between dielectric plates demonstrates
that there is an ample evidence of problems between the Lifshitz theory using the
dielectric response function for real dielectrics and
thermodynamics, on the one hand,
and measurement data, on the other hand. Experiments on measuring
the Casimir interaction with dielectric test bodies are more complicated in
comparison with metallic ones due to the problem of localized electric charges which
are present on dielectric surfaces.  Up to now, only three relatively precise
measurements using the dielectric test bodies have been performed. However, the
thermodynamic test similar to that elucidated in Section~4.1 for metals is even more
conclusive because it does not require an assumption of the perfect crystal lattice.
We begin with the thermodynamic argument and continue with the comparison between
experiment and theory.


\subsection{The Casimir Entropy for Dielectric Plates and the Nernst Heat Theorem}

The thermodynamic test for two dielectric plates with either included or omitted
dc conductivity can be performed in close analogy to that presented in Section~4.1
for metallic plates described either by the Drude or by the plasma model.
One should substitute the dielectric permittivity of a real dielectric (\ref{eq39})
with $\ve(\ix_l)$ given by (\ref{eq34}) in \\(\ref{eq7}) in order to obtain the
Casimir free energy ${\cal F}_{\rm RD}$ for real dielectric plates and then find the
low-temperature behavior of the Casimir entropy
\begin{equation}
S_{\rm RD}(a,T)=-\frac{\partial{\cal F}_{\rm RD}(a,T)}{\partial T}.
\label{eq43}
\end{equation}
\noindent
In a similar way, substituting the dielectric permittivity  (\ref{eq34}) in (\ref{eq7}),
one obtains the Casimir free energy ${\cal F}_{\rm ID}$ for ideal dielectric plates
with omitted dc conductivity and then, like in (\ref{eq43}), finds their Casimir
entropy $S_{\rm ID}$.

The Casimir entropy $S_{\rm RD}$ can be presented in the form \cite{9,30,31}
\begin{eqnarray}
&&
S_{\rm RD}(a,T)=S_{\rm ID}(a,T)+\frac{k_B}{4\pi}\int_{0}^{\infty}\!\!\kb d\kb
\left\{\ln\left[1-r_{\rm TM}^2(0,\kb)e^{-2a\kb}\right]\right.
\nonumber \\
&&~~~~~~~~~~~~~~~~~~~~~~~\left.
-\ln\left(1-e^{-2a\kb}\right)\right\}
+O\left(\frac{1}{T^3}e^{-\frac{\Delta}{k_BT}}\right),
\label{eq44}
\end{eqnarray}
\noindent
where the reflection coefficient $r_{\rm TM}(0,\kb)$ defined in (\ref{eq36}) does
not depend on $\kb$: $r_{\rm TM}(0,\kb)=r_{\rm TM}(0)$. Because of this, the
integral in (\ref{eq44}) can be calculated with the result
\begin{eqnarray}
&&
S_{\rm RD}(a,T)=S_{\rm ID}(a,T)+\frac{k_B}{16\pi a^2}
\left\{\zeta(3)-Li_3\left[r_{\rm TM}^2(0)\right]\right\}
\nonumber \\
&&~~~~~~~~~~~~~~~~~~~~~~~~~~~~~~~~~~~
+O\left(\frac{1}{T^3}e^{-\frac{\Delta}{k_BT}}\right).
\label{eq45}
\end{eqnarray}

The asymptotic behavior of the Casimir entropy for an ideal dielectric at arbitrary low
temperature is given by \cite{9,30,31}
\begin{equation}
S_{\rm ID}(a,T)=\frac{k_B^2T}{2\hbar c a^2}\left\{
\frac{G Li_3\left[r_{\rm TM}^2(0)\right]}{3[\ve^2(0)-1]}+
\frac{3\zeta(3)r_{\rm TM}^2(0)[\ve^2(0)+1]}{2\pi}\,\frac{a^2k_BT}{\hbar c}
+O\left(\frac{T}{T_{\rm eff}}\right)^2\right\},
\label{eq46}
\end{equation}
\noindent
where the quantity $G$ is expressed via the oscillator parameters introduced in (\ref{eq34})
\begin{equation}
G=c\sum_{j=1}^{K}\frac{g_j\gamma_j}{\omega_j^4}.
\label{eq47}
\end{equation}

As is seen from (\ref{eq46}),
\begin{equation}
\lim_{T\to 0}S_{\rm ID}(a,T)=S_{\rm ID}(a,0)=0,
\label{eq48}
\end{equation}
\noindent
i.e., the Lifshitz theory satisfies the Nernst heat theorem in the case of Casimir
plates made of ideal dielectric. In this sense the ideal dielectric plates are similar
to the ideal metal plates and to the plates made of a metal described by the plasma model.

The situation reverses when we consider the Casimir entropy of real dielectric plates
made of real dielectric possessing some dc conductivity at any nonzero temperature.
In the limiting case of zero temperature (\ref{eq46}) results in
\begin{equation}
\lim_{T\to 0}S_{\rm RD}(a,T)=S_{\rm RD}(a,0)=\frac{k_B}{16\pi a^2}
\left\{\zeta(3)-Li_3\left[r_{\rm TM}^2(0)\right]\right\}>0,
\label{eq49}
\end{equation}
\noindent
where $r_{\rm TM}(0)$ is given in (\ref{eq36}). Thus, for the plates made of real
dielectric possessing some dc conductivity at any nonzero temperature the Lifshitz
theory leads to a positive Casimir entropy depending on the separation between the
plates and the static dielectric permittivity, i.e., violates the Nernst heat
theorem. Note that for both metallic and dielectric plates the Casimir entropy is not
the entropy of a closed system which should also include, for instance, the
entropy of plate materials. It should be stressed, however, that the Casimir
entropy alone depends on the separation distance between the plates. Because of this
it cannot be compensated by some other contribution in the total entropy of a
closed system making the violation of the Nernst heat theorem unavoidable.

The presented results demonstrating that with taken into account dc conductivity
of plate material the Lifshitz theory violates the Nernst heat theorem and satisfies
it when the dc conductivity is disregarded are called {\it the Casimir conundrum}.
These results are really strange if to take into account that the dc conductivity
does exist and has been measured in numerous experiments.

\subsection{The Lifshitz Theory Confronts Experiments with Dielectric Test Bodies}

As mentioned in the beginning of Section~6, measurements of the Casimir force
between dielectric test bodies are more complicated than between metallic ones
due to the localized electric charges which are present on dielectric surfaces.
Because of this, it is preferable to use dielectrics with sufficiently high
concentration of charge carries (for instance, doped semiconductors with doping
concentration below the critical value at which the dielectric-to-metal phase
transition occurs).

We start with an experiment on the optically modulated Casimir force where the
force difference between an Au-coated sphere of $R=98.9~\mu$m radius and a silicon
membrane illuminated with laser pulses has been measured in high vacuum in the
presence and in the absence of light \cite{36,37}. Measurements of the force
difference
\begin{equation}
F_{\rm diff}(a,T)=F_{sp}^l(a,T)-F_{sp}(a,T)
\label{eq50}
\end{equation}
\noindent
have been performed by means of an atomic force microscope.

In the absence of a laser pulse on the membrane made of a p-type silicon the
concentration of charge carriers in it was equal to approximately
$5\times 10^{14}~\mbox{cm}^{-3}$ \cite{50}. Thus, the membrane was in a
dielectric state. In the presence of a laser pulse the charge carrier
concentration increased by the five orders of magnitude up to
$(1-2)\times 10^{19}~\mbox{cm}^{-3}$ depending on the absorbed power.
This means that in the bright phase silicon was in a metallic state.

\begin{figure}[!t]
\centering
\vspace*{-6.5cm}
\hspace*{-1.6cm}\includegraphics[width=16 cm]{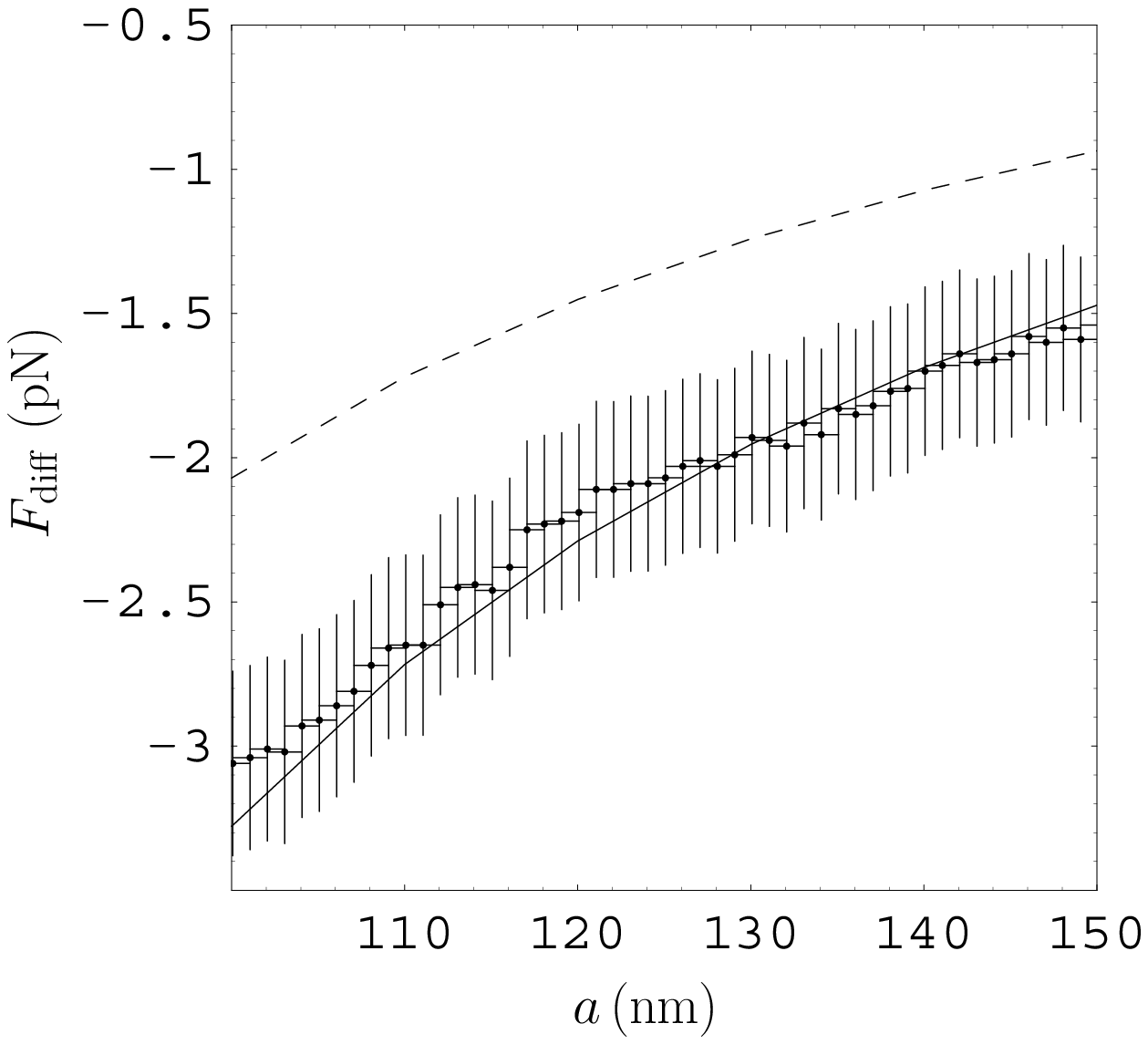}
\vspace*{-7.5cm}
\caption{The difference of the Casimir forces between an Au-coated sphere
and a silicon membrane in the metallic and dielectric states measured by means
of an atomic force microscope \cite{37} (crosses) are compared with
theoretical predictions of the Lifshitz theory obtained with omitted and
included dc conductivity of Si in the dielectric state (solid and dashed
lines, respectively).
\label{fg5}}
\end{figure}

The difference of the Casimir forces between an Au-coated sphere and a silicon
membrane in the bright and dark phases was calculated using the Lifshitz theory
and the proximity force approximation (\ref{eq32}) with account of surface
roughness by means of the geometrical averaging at the laboratory temperature
$T=300~$K. In Figure~\ref{fg5} the computational results for $F_{\rm diff}$
are shown by the solid and dashed lines as the functions of sphere-plate
separation (the absorbed power was 8.5~mW). The solid and dashed lines were
computed with omitted and taken into account dc conductivity of silicon,
respectively, in the absence of a laser pulse, i.e., in the dark phase.
In the presence of a laser pulse (the bright phase) the charge carriers were
taken into account by means of the Drude model (the use of the plasma model
in the bright phase for silicon and for Au leads to only minor differences in
this case which cannot be distinguished experimentally). The experimental
data are shown as crosses plotted at the 95\% confidence level.

As is seen in Figure~\ref{fg5}, the predictions of the Lifshitz theory taking
into account the dc conductivity of silicon in the dielectric state (the
dashed line) are experimentally excluded whereas the predictions with omitted
dc conductivity of dielectric silicon are in a very good agreement with the
measurement data. Thus, the Lifshitz theory is experimentally consistent only
if one ignores the real physical phenomenon --- small but quite measurable electric
conductivity. It is highly meaningful that just in this case the theory is in
agreement with the requirements of thermodynamics and is in contradiction with them
otherwise.

One more experiment performed by using an atomic force microscope was devoted to
measurements of the Casimir force between an Au-coated sphere of $R=101.2~\mu$m
radius and an indium tin oxide (ITO) film deposited on a quartz substrate.
These measurements have been performed for two times --- before and after a UV
irradiation of the ITO film \cite{39,40}. At room temperature ITO is a transparent
conductor. Based on this, it was suggested to use this material in measurements
of the Casimir force \cite{13}. First experiments of this kind have been performed
earlier \cite{65,66}. The main novelty of the experiments \cite{39,40} is a UV
irradiation of the ITO film with subsequent measurement of the Casimir force.
The point is that the UV irradiation of ITO leads to a lower mobility of charge
carriers \cite{66a} with no sensible changes in the optical data  and,
as was hypothesized in \cite{39,40}, to a phase transition from a metallic to
a dielectric state.

\begin{figure}[!t]
\centering
\vspace*{-14.cm}
\hspace*{-5.5cm}\includegraphics[width=32 cm]{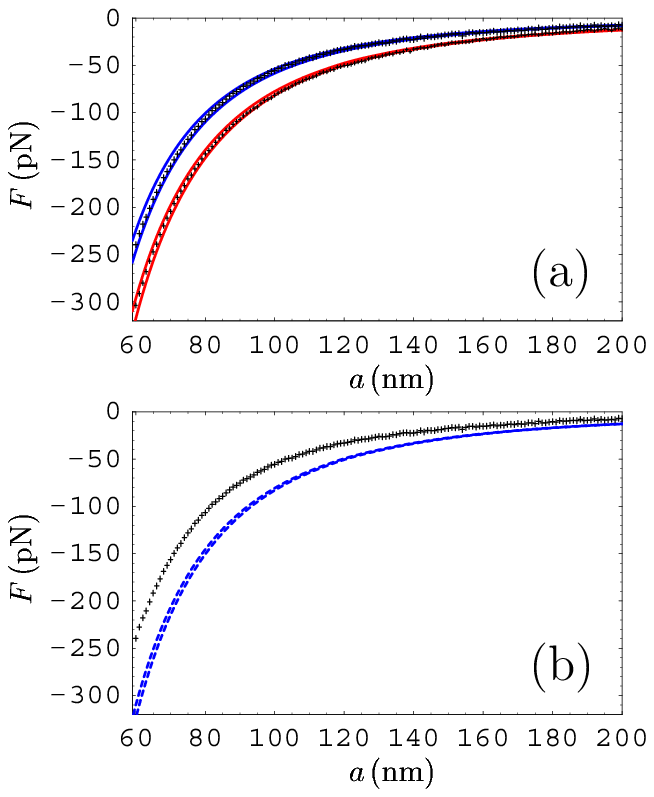}
\vspace*{-18.6cm}
\caption{The  Casimir forces between an Au-coated sphere
and an ITO-coated plate  measured by means of an atomic force microscope are shown
as crosses \cite{39,40}. (a) The lower and upper sets of crosses represent the
measurement results before and after a UV irradiation of the ITO film.
The red and blue theoretical bands are computed with taken into account and disregarded
free charge carriers, respectively.
(b)The set of crosses represent the measurement results after a UV irradiation of the
ITO film. The blue theoretical band is computed with taken into account free charge
carriers in ITO.
\label{fg6}}
\end{figure}

The experimental results and their comparison with theory confirmed this hypothesis.
In Figure~\ref{fg6}(a), the pairs of blue and red lines indicate the boundaries of
theoretical bands for the Casimir force between an Au sphere and an ITO film
computed before and after a UV irradiation of this film with taken into account and
omitted contribution of free charge carriers, respectively. The experimental data
in both cases are shown as crosses plotted at the 95\% confidence level.
It is seen that the theoretical predictions are in a very good agreement with the
measurement data both before and after irradiation.

In Figure~\ref{fg6}(b) we again plot the upper set of crosses obtained after
a UV irradiation of the ITO film in comparison with the theoretical band
computed with taken into account contribution of free charge carriers in the
UV irradiated sample. It is seen that the predictions of the Lifshitz theory
are excluded by the measurement data. This result is in line with the results
found from measuring the optically modulated Casimir force which show that
in the dielectric state an agreement between experiment and theory is reached by
disregarding the contribution of free charge carriers to the dielectric response
of a material.

The last experiment utilizing the dielectric test body was on measuring the
thermal Casimir-Polder force between ${}^{87}$Rb atoms belonging to the
Bose-Einstein condensate and a dielectric fused silica plate at separations of
a few micrometers \cite{41}. In the first series of measurements the plate was
kept at the same temperature as the environment and in the second and third at
higher temperatures. This means that in the last two cases the Casimir-Polder
force had two contributions --- the equilibrium one given by the Lifshitz
theory and another one from its generalization for a nonequilibrium case
\cite{67,68} (see also the case of phase-change wall material at
nonequilibrium conditions \cite{69}). It was shown \cite{41} that the measurement
data are in good agreement with theory when the dc conductivity of fused silica
is not taken into account in computations. However, by repeating the same
calculations with included dc conductivity of the plate material, one obtains
the theoretical results excluded by the measurement data \cite{38}.
This happens due to exclusively the equilibrium contribution to the Casimir-Polder
force given by the Lifshitz theory (the nonequilibrium one is almost independnet
on whether or not the dc conductivity is taken into account in computations).

Thus, experiments on measuring the Casimir interaction with dielectric test bodies
are consistent with theoretical predictions of the Lifshitz theory only under
a condition that the dc conductivity of dielectric material is omitted in
computations. In so doing, as shown in Section~6.1, the theory is consistent with
the laws of thermodynamics, but at the sacrifice of the phenomenon of dc
conductivity. The approaches to resolution of the Casimir conundrum originating
from this situation are discussed below.

\section{Different Approaches to Resolution of the Casimir Puzzle and
Casimir Conundrum}

The above problems attracted much attention of experts in the field during the last
20 years. A lot of different explanations and suggestions were put forward on why
the predictions of the Lifshitz theory appear to be in disagreement with the laws
of thermodynamics and the measurement data and how the agreement could be restored.
Certain of the proposed approaches were seeking for some unaccounted systematic
effects in the performed experiments, such as, for instance, an impact of the additional
electric force due to electrostatic patches on metallic surfaces or the nonadditive
effects in the surface roughness. According to other approaches, the roots of the
problems could lay in simplifications used in the Lifshitz theory which does not take
into account the spatial dispersion.
There even was a suggestion \cite{69a} to modify the Planck distribution by including
in it the special damping parameter. This parameter, however, should take different values
for bringing theoretical predictions in agreement with the measurement data of different
experiments.
The separate line of investigation was a
generalization of the Lifshitz theory for more complicated geometries used in
experiments, including the configuration of a sphere above a plate, basing on the
first principles of quantum field theory with no use of uncontrolled additive
methods such as PFA. Below we consider several suggested approaches directed to the
elimination of contradictions between the Lifshitz theory, on the one hand,
and thermodynamics and/or experiment, on the other hand.

\subsection{Variations of the Optical Data}

Many precise experiments on measuring the Casimir force between metallic test bodies
used the Au-coated surfaces of a sphere and a plate. The frequency-dependent dielectric
permittivity of Au was found using the tabulated optical data for the complex index of
refraction \cite{50} extrapolated down to zero frequency by means of either the Drude
or the plasma models as discussed above. It has been known, however, that the
optical properties of Au film are sample-dependent and also depend on the used
method of deposition. Most importantly, different sets of the optical data lead to
different values of the Drude parameters $\omega_p$ and $\gamma$ used in extrapolation
to zero frequency. This may influence on the values of the Casimir force calculated
using the Lifshitz theory and on the comparison between experiment and theory.
The question arises whether it is possible to bring the theoretical predictions in
agreement with the experimental results if the optical data are not taken from the tables but
measured for the laboratory bodies used in this specific experiment \cite{70,71}.
To answer this question, the Casimir force was computed using different sets of the
optical data for Au available in the literature. It was concluded \cite{70,71} that
the variations in the optical data of Au with respective changes in the plasma
frequency used in extrapolations to zero frequency may result in up to 5--10\%
differences in the Casimir force. These differences are of the same size as the
discrepancies between experiment and theory found in \cite{19,20,21,22,23,24,25}.

\begin{figure}[!t]
\centering
\vspace*{-2.2cm}
\hspace*{-1.6cm}\includegraphics[width=16 cm]{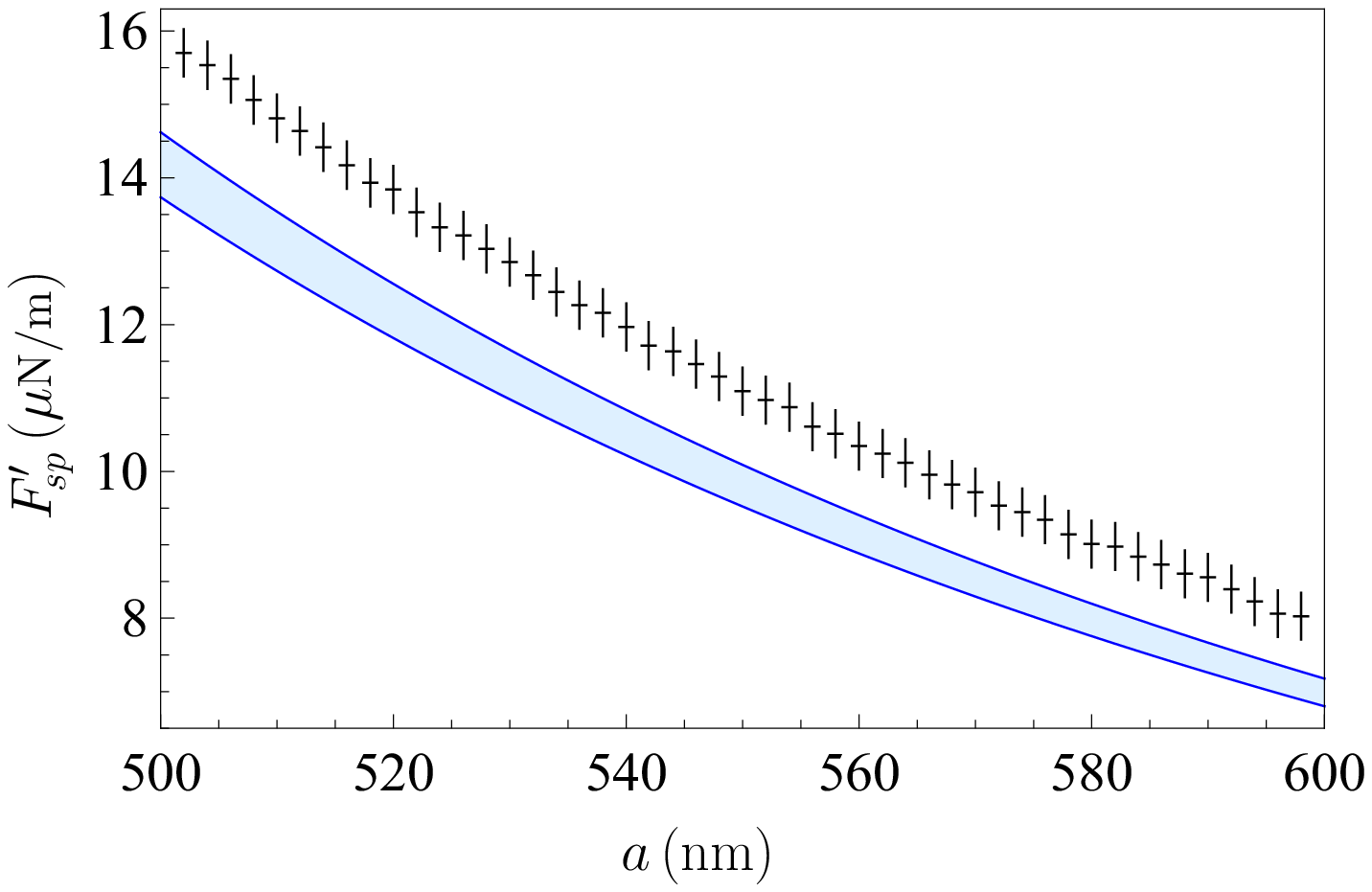}
\vspace*{-13.cm}
\caption{The gradients of the Casimir force between an Au-coated sphere and
an Au-coated plate measured by means of micromechanical torsional oscillator
\cite{21} (crosses) are compared with theoretical predictions of the Lifshitz
theory using all available sets of the optical data for the complex index of
refraction of Au extrapolated to zero frequency by the Drude  model
(blue band).
\label{fg7}}
\end{figure}
To determine the actual role of the optical data variations in resolution of the
Casimir puzzle in the experiment \cite{22}, the plasma frequency and the relaxation
parameter were found for the specific Au coatings used in measurements of the
Casimir force. The obtained values turned out to be very close to the tabulated
ones \cite{50}. In addition, a complete set of the optical data for experimental Au
coatings  was measured by means of ellipsometry and found \cite{72} in agreement
with the tabulated one. These results were confirmed by the method of the weighted
Kramers-Kronig relations \cite{73}. Moreover, all sets of the optical data of Au
available in the literature with respective plasma frequencies varying from 6.85
to 9.0~eV have been used in calculations of the Casimir interaction for subsequent
comparison with the measured values of \cite{22}. The results of this comparison
are shown in Figure~\ref{fg7} where the measured gradients of the Casimir force
are shown as crosses (the same as in Figure~\ref{fg2}) and the blue theoretical
band is obtained using all sets of the optical data of Au extrapolated to zero
frequency by means of the Drude model. From Figure~\ref{fg7} it is seen that the
use of alternative sets of the optical data only increases a disagreement between
experiment and theory. Thus, it was concluded \cite{55} that it is not possible to
bring the predictions of the Lifshitz theory using the Drude model in agreement with
the measurement data at the cost of variation of the optical data.

\subsection{Impact of the Surface Patches}

The Au coatings on the test bodies in experiments on measuring the Casimir force
have a polycrystal structure. In addition, even in high vacuum, there are some
contaminants and dust on the surfaces. As a result, some spatial distribution of
electrostatic potentials appear even on the grounded surfaces of a sphere and a
plate which are called {\it the patch potentials} \cite{74}. The patch potentials
lead to some additional attractive force acting between a sphere and a plate
and to a dependence on separation of the residual potential difference between them.
The magnitude of this force depends on the size of surface patches or contaminants.

It was hypothesized \cite{75} that an additional force gradient due to surface
patches could compensate a difference between the blue and red bands in Figure~\ref{fg2}
and, thus, bring the Lifshitz theory using the Drude model in agreement with the measurement
data of \cite{21}. Sufficiently large patches and contaminants up to $2~\mu$m size can really
lead to the force gradient of required magnitude but also result in strong dependence of the
residual potential difference on the sphere-plate separation which was not present in
the calibration measurements of \cite{19,20,21,22,23,24,25} due to a special selection of
samples. An investigation \cite{75a} of the patch potentials on Au surfaces of the samples
similar to those used in the performed experiments \cite{21,22} by means of Kelvin probe
force microscopy demonstrated that the force of patch origin cannot explain a difference
between the measured force gradients and theoretical predictions of the Lifshitz
theory using the Drude model.

Further clarity to the role of patches in measurements of the Casimir force was added
by the experiments with magnetic test bodies \cite{24,25}. As noted in Section~4.2,
for bringing the theoretical predictions obtained using the Drude model in agreement
with the measurement data in this case, one needs some additional {\it repulsive}
force which does not occur due to surface patches. It should be mentioned also that
recently several experiments on measuring the Casimir interaction have been performed
where the role of surface patches and contaminants was strongly suppressed by means
of Ar-ion and UV-cleaning of the interacting surfaces \cite{76,77,78}.
The results of these experiments are consistent with theoretical predictions of the
Lifshitz theory using the plasma model and exclude the predictions of the same theory
using the Drude model up to the separation distance of $1.1~\mu$m. Thus, by accounting
the surface patches it is impossible to reconcile the Lifshitz theory using the Drude
model with the measurement data.

\subsection{The Role of Surface Roughness}

The interacting surfaces in measurements of the Casimir force are not perfectly plane
or spherical, but covered with some roughness. If there is a large-scale deviations
from the perfect geometry, it can be described by a regular function. The short-scale
roughness can be considered as stochastic and described by the random functions.
In the comparison between precise experiments on measuring the Casimir force mentioned
above and theory the roughness was taken into account in an additive way \cite{9,55,57}.
This means that in the case of a regular roughness the Casimir force was computed
using the Lifshitz theory at all local separations and was then averaged over the actual
geometrical profiles of interacting surfaces determined by means of an atomic force
microscope. In the case of stochastic roughness with small dispersions
$\delta_{1,2}\ll a$ the roughness correction can be found perturbatively in powers of
$\delta_{1,2}/a$ and taken into account in a multiplicative way. For example,
a corrected for the presence of stochastic roughness gradient of the Casimir force
between a sphere and a plate takes the form \cite{9,55}
\begin{equation}
F_{sp,R}^{\prime}(a,T)= F_{sp}^{\prime}(a,T)\left[1+
10\frac{\delta_1^2+\delta_2^2}{a^2}+
105\frac{(\delta_1^2+\delta_2^2)^2}{a^4}\right].
\label{eq51}
\end{equation}

Different approaches to the account of surface roughness in experiments on measuring
the Casimir force are the subject of considerable literature \cite{79,80,81,82,83}.
One may doubt whether an additive approach used in comparison between experiment and
theory is exact enough. Strictly speaking, the larger and smaller roughness corrections
to the Casimir force could decrease a difference between the predictions of the Lifshitz
theory using the Drude model and the measurement data in experiments with the
nonmagnetic and magnetic surfaces, respectively.

The more fundamental scattering approach to the Casimir force between rough surfaces
was developed in \cite{84,85}. It was shown that the additive approach leads to
sufficiently precise results at much smaller separations between the interacting bodies
as compared with the roughness correlation length.  In fact for a typical correlation
length $\Lambda_c\approx 200~$nm the additive approach is already applicable at
$a<2\Lambda_c/3\approx 130~$nm. This is the most important region where the roughness
correction contributes up to a few percent fraction of the Casimir force.
At larger separations the additive approach underestimates the role of surface
roughness, so that the more accurate results are given by the scattering approach.
For instance, at $a=2\Lambda_c=400~$nm the scattering approach leads to by a factor
1.5 larger roughness correction than the additive one. This is, however, not
damaging for the comparison between experiment and theory. The point is that with
increasing separation the roughness correction decreases much faster  than the
Casimir force between perfectly shaped surfaces. Because of this for surfaces of
sufficiently good quality at separations of a few hundred nanometers it does not play
any role in the comparison between experiment and theory and can be simply neglected.
This means that the surface roughness is not helpful for resolution of the Casimir
puzzle.

\subsection{Deviations from the Proximity Force Approximation}

As was mentioned in Section~4.2, calculations of the Casimir force and its gradient
in the sphere-plate geometry used in comparison between experiment and theory were
performed by means of the Lifshitz theory and the proximity force approximation
(\ref{eq32}) and (\ref{eq33}). Although it was believed that for small  $a/R$ the
error introduced by using the PFA should be of the order of $a/R$, the exact
information concerning the size of this error was missing. Because of this one could
hope that a disagreement between theoretical predictions of the Lifshitz theory using
the Drude model and the measurement data may be caused by some deviations from the
PFA in the region of experimental separations and sphere radii.

A major step forward was made in 2006 when it was shown that the Casimir energy
between nonplanar surfaces can be written in terms of functional determinants
\cite{86,87}. The obtained results were applied to calculate the Casimir energy
of an ideal metal cylinder and a sphere in front of an ideal metal plane
\cite{86,87,88}. Later the Lifshitz theory was generalized for real material
bodies of arbitrary shape kept at any temperature in thermal equilibrium with the
environment \cite{89,90,91,92}. Application of the developed theory to a sphere
and a plate made of real metals spaced at separations below a micrometer turned
out to be a complicated problem.  First computations \cite{93} produced an
impression that in the limiting case of perfect reflectors the exact results
rapidly depart from the PFA expectations and this fact should be taken into
account in theory-experiment comparison. For real metals the computations
have been performed in \cite{93a} but for not too small values of $a/R>0.2$, as
compared to the experimental values  $a/R<0.01$.

Calculations of the Casimir force in a sphere-plate geometry for the experimental
values of $a/R$ have been performed both analytically using the method of gradient
expansion and numerically \cite{94,95,96,97,98,99,100,101}. As only one example
of the obtained result, we present an expansion of the exact gradient of the
Casimir force between a sphere and a plate which includes the first-order
correction to the PFA \cite{99}
\begin{equation}
F_{sp,{\rm ex}}^{\prime}(a,T)=F_{sp}^{\prime}(a,T)\left[1+\beta_{D,p}(a,R)
\frac{a}{R}\right],
\label{eq52}
\end{equation}
\noindent
where $F_{sp}^{\prime}$ is the PFA result presented in (\ref{eq33}).
The values of the expansion coefficient $\beta_{D,p}$ were computed at different
separations for the sphere radii varying from 10 to $100~\mu$m using the tabulated
optical data for the complex index of refraction of Au extrapolated to zero
frequency by the Drude ($\beta_D$) and plasma ($\beta_p$) models.
According to the obtained results \cite{99}, within the region of separations from
200 to 600~nm the coefficient $\beta_D$ varies from --0.35 to --0.45 whereas
$\beta_p$ varies from --0.5 to --0.6. Note that in the interpretation of
experiments on measuring the Casimir force \cite{19,20,21,22,23,24,25} the
conservative estimation $|\beta_{D,p}|\leqslant 1$ has been used.
Thus, inaccuracies in the PFA are irrelevant to the Casimir puzzle and an
exact computation of the Casimir force in the sphere-plate geometry does not
help to resolve it.

\subsection{Impurities in a Crystal Lattice and the Nernst Heat Theorem}

The above consideration of the variations of the optical data, impact of surface
patches, nonadditive effects in the surface roughness, and deviations of the
Casimir force from PFA is aimed at resolving the second, experimental, part
of the Casimir puzzle. All these research directions are unrelated to an
inconsistency of the Lifshitz theory with thermodynamics discussed in
Sections~4.1 and 6.1 (note that in the sphere-plate geometry the Lifshitz
theory faces the same problems with violation of the Nernst heat theorem as
for two parallel plates \cite{15}). It should be stressed also that after the
crucial experiment \cite{26}, where the theoretical predictions using the plasma
and Drude models differed by up to a factor of 1000, an exclusion of the latter
by the measurement data was established conclusively. This attaches particular
significance to the thermodynamic parts of the Casimir puzzle and Casimir
conundrum.

The historically first approach to the problem of violation of the Nernst heat
theorem in the configuration of two parallel metallic plates was made with no
modifications in the Lifshitz theory. It was suggested \cite{102} to take into
account that the crystal lattice of any real metal has some small fraction of
impurities. As a result, the relaxation parameter $\gamma$ of the Drude model
(\ref{eq16}) does not go to zero with vanishing temperature but to some residual
value $\gamma_0$ which depends on the impurity concentration \cite{53}.
Based on this, it was shown numerically \cite{102} that at sufficiently low
temperature the Casimir entropy abruptly jumps to zero starting from the
negative value (\ref{eq31}), i.e., the Nernst heat theorem is formally satisfied.

It should be taken into account, however, that the numerical proof of vanishing
cannot be considered as completely satisfactory because it is burdened with some
computational error. Because of this, the case of metallic plates with impurities
was also considered analytically \cite{103,104}. As a result, it was shown that
in the asymptotic limit of low $T$ the Casimir entropy calculated using the Drude
model with the residual relaxation $\gamma_0$ vanishes as \cite{104}
\begin{equation}
S_D(a,T)=-D_1(a)T+D_2(a)T^{3/2}-\,\ldots \,.
\label{eq53}
\end{equation}
\noindent
For the typical value of  $\gamma_0=5.32\times 10^{10}~$rad/s, one finds from
(\ref{eq53}) that the Casimir entropy jumps to zero at about 0.001~K starting
from the negative value of $-2\mbox{MeV\,m}^{-2}\mbox{K}^{-1}$ \cite{9}.

Although an account of impurities provides an apparent resolution of the
thermodynamic part of the Casimir puzzle, it cannot be considered as completely
satisfactory. The point is that the perfect crystal lattice with no impurities
serves as the basis for quantum condensed matter physics. It possesses the
nondegenerate ground state and the Nernst heat theorem must be satisfied in this
case as well \cite{105,106}. In Section~8, we will return to this point in
connection with the recently proposed Drude-like nonlocal response functions.

As discussed in Section~6.1, the thermodynamic part of the Casimir conundrum for
dielectrics is due to the exponentially fast vanishing of the conductivity
(\ref{eq40}) with decreasing temperature. This leads to a violation of the Nernst
heat theorem if the conductivity is taken into account in the Lifshitz theory.
To avoid this conclusion in a similar way as for metals with impurities, it was
suggested \cite{107} to consider some imaginary dielectric material possessing a
constant conductivity at low temperature. Although the proposed model does not
carry the thermodynamic anomaly, it is incapable to solve the Casimir conundrum
for real dielectric materials because they are characterized by the exponentially
fast vanishing conductivity at low temperatures. Thus, neither the Casimir
puzzle nor the Casimir conundrum can be solved without some radical changes
in the Lifshitz theory and/or in the used response functions.

\subsection{The Anomalous Skin Effect and Spatial Nonlocality}

The original formulation of the Lifshitz theory assumes that the material of the
plates is described by the dielectric permittivity depending on the frequency but
not on the wave vector. In the case of metallic plates there is, however, the
frequency region where a connection between the electric field and the current
becomes nonlocal and the concept of the frequency-dependent dielectric permittivity
loses its meaning. This is the region of the anomalous skin effect where the
spatial nonlocality plays an important role. At room temperature for Au the frequency
region of the anomalous skin effect is from $10^{12}$ to $10^{13}~$rad/s,
i.e., it is rather narrow. With decreasing temperature, however, this frequency
region widens at the cost of a decrease of its left boundary. The question arises
what is an impact of the anomalous skin effect on the Casimir force and what is its
possible role in the resolution of the Casimir puzzle.

In the presence of spatial nonlocality the reflection coefficients in the Lifshitz
formulas (\ref{eq7}) and (\ref{eq9}) are expressed via the surface impedances for
the TM and TE polarizations \cite{108,109,110}
\begin{equation}
r_{\rm TM}(\ix_l,\kb)=
\frac{cq_l-\xi_lZ_{\rm TM}(\ix_l,\kb)}{cq_l+\xi_lZ_{\rm TM}(\ix_l,\kb)},
\qquad
r_{\rm TE}(\ix_l,\kb)=
\frac{cq_lZ_{\rm TE}(\ix_l,\kb)-\xi_l}{cq_lZ_{\rm TE}(\ix_l,\kb)+\xi_l}.
\label{eq54}
\end{equation}
\noindent
The impedances in turn are expressed via the longitudinal
$\ve^L(\ix_l,\mbox{\boldmath$k$})$ and transverse
$\ve^T(\ix_l,\mbox{\boldmath$k$})$ dielectric permittivities which describe
a dielectric response of metal to parallel and perpendicular to {\boldmath$k$}
electric fields, respectively \cite{109,110}
\begin{eqnarray}
&&
Z_{\rm TM}(\ix_l,\kb)=\frac{\xi_l}{\pi c}\int_{-\infty}^{\infty}\!
\frac{dk_z}{k^2}\left(\frac{c^2k_{\bot}^2}{\xi_l^2\ve^L(\ix_l,\mbox{\boldmath$k$})}
+\frac{k_z^2}{{k^T}^2(\ix_l,\mbox{\boldmath$k$})+k_z^2}\right),
\nonumber \\
&&
Z_{\rm TE}(\ix_l,\kb)=\frac{\xi_l}{\pi c}\int_{-\infty}^{\infty}\!
\frac{dk_z} {{k^T}^2(\ix_l,\mbox{\boldmath$k$})+k_z^2}.
\label{eq55}
\end{eqnarray}
\noindent
Here, $k_z$ is the third component of the wave vector, so that $k^2=k_{\bot}^2+k_z^2$
and
\begin{equation}
{k^T}^2(\ix_l,\mbox{\boldmath$k$})=k_{\bot}^2+\ve^T(\ix_l,\mbox{\boldmath$k$})
\frac{\xi_l^2}{c^2}.
\label{eq56}
\end{equation}

The nonlocal generalizations $\ve^{L,T}$ of the Drude dielectric function $\ve_D$
for the description of the anomalous skin effect suggested in \cite{109} were
used in \cite{112} to calculate the correction to the Casimir force due to the
effect of nonlocality. It was shown that with increasing separation from 100 to
300~nm the relative nonlocal correction is negative and its magnitude decreases
from 0.3\% to 0.1\%. Thus, an account of the effects of nonlocality arising in
the dielectric response due to the anomalous skin effect cannot solve the
experimental part of the Casimir puzzle.

Concerning the thermodynamic part of the Casimir puzzle, an account of the
anomalous skin effect leads to the same conclusions as were discussed above for the simple
Drude model. The Nernst heat theorem is formally restored if there is some nonzero value of
the effective relaxation parameter at zero temperature and is violated otherwise
\cite{113}. In this sense the spatial dispersion can play the same role as the
relaxation of conduction electrons \cite{113a}.

\subsection{Inclusion of the Screening Effects}

One more attempt to solve the Casimir puzzle and Casimir conundrum was made by
taking into account the screening of the electric field created by an external sourse
in a medium containing free charge carriers. This effect leads to a penetration
of the static electric field into a conducting material to a depth of the so-called
{\it screening length} \cite{114}.

In  the Casimir physics an account of the screening effects was first proposed in \cite{115} in
connection with an experiment on measuring the Casimir-Polder force between a ${}^{87}$Rb
atom and a dielectric fused silica plate \cite{41} (see Section~6.2). If the dc
conductivity of fused silica is taken into account in calculations of the Casimir-Polder
force, the theoretical predictions of the standard Lifshitz theory are excluded by the measurement
data. The Casimir-Polder entropy at zero temperature is equal to
\begin{equation}
S(a,0)=\frac{k_B\alpha(0)}{4a^3}[1-r_{\rm TM}(0)]>0,
\label{eq57}
\end{equation}
\noindent
where $r_{\rm TM}(0)$ is defined in (\ref{eq36}) and $\alpha(0)$ is the static
polarizability  of ${}^{87}$Rb atom, in violation of the Nernst heat theorem \cite{38}.

With account of screening, the TM reflection coefficient at zero frequency (\ref{eq36})
is replaced with \cite{115}
\begin{equation}
r_{\rm TM}^{\rm src}(0,\kb)=
\frac{\ve(0)\sqrt{\kappa^2+k_{\bot}^2}-\kb}{\ve(0)\sqrt{\kappa^2+k_{\bot}^2}+\kb},
\label{eq58}
\end{equation}
\noindent
where $\kappa$ is the inverse quantity to the Debye-H\"{u}ckel screening length
\begin{equation}
\frac{1}{\kappa}\equiv\frac{1}{\kappa_{DH}}=\sqrt{
\frac{\ve_0k_BT}{4\pi e^2n}}
\label{eq59}
\end{equation}
\noindent
and $n=n(T)$ is the concentration of charge carriers in a dielectric material.

According to the results of \cite{115}, the Lifshitz theory using the screened
reflection coefficient (\ref{eq58}) taking into account the presence of free charge
carriers is consistent with the measurement data. This solves the experimental part
of the Casimir conundrum in application to this specific experiment. It was shown
\cite{116}, however, that the same theory still remains in disagreement with the
measurement data of an experiment on the optically modulated Casimir force
(see Section~6.2).

The Lifshitz theory with the screened
reflection coefficient (\ref{eq58}) also provides a partial resolution of the
thermodynamic part of the Casimir conundrum. It turns out that for the insulators
and intrinsic semiconductors, whose concentration of charge carriers $n(T)$ vanishes
with temperature exponentially fast, the Nernst heat theorem is satisfied.
There are, however, dielectric materials (dielectric-type semimetals, doped
semiconductors with the dopant concentration below critical, dielectrics with ionic
conductivity, etc) whose static conductivity $\sigma(0)$ vanishes with temperature not
due to the vanishing $n$ but due to the vanishing mobility of charge carriers.
For these dielectric materials the Lifshitz theory using  the screened
reflection coefficient (\ref{eq58}) still leads to a violation of the Nernst heat
theorem \cite{116}.

In the end of this section, we note that the screening effects were also taken into
account in the reflection coefficients with any $l$. The developed formalism was
presented in the form applicable to bodies with arbitrarily large charge carrier
density including metals \cite{117}. For this purpose the Debye-H\"{u}ckel screening
length should be replaced with the Thomas-Fermi one \cite{114}. It was shown \cite{116},
however, that this approach leads to approximately the same Casimir forces between
metallic plates, as given by the standard Lifshitz theory using the Drude model, and,
thus, remains to be in contradiction with the measurement data. For metals with perfect
crystal lattices the modified Lifshitz theory accounting for the screening effects
violates the Nernst heat theorem. Thus, in spite of some encouraging results, an account
of the screening effects did not resolve the Casimir puzzle and Casimir
conundrum.

\newcommand{\veT}{{\varepsilon^T({\rm i}\xi_l,k_{\bot})}}
\newcommand{\veL}{{\varepsilon^L({\rm i}\xi_l,k_{\bot})}}
\newcommand{\tve}{\tilde{\varepsilon}}

\section{The nonlocal Drude-Like Response to Quantum Fluctuations off the Mass
Shell and the Casimir Puzzle}

Many unsuccessful attempts to solve the Casimir puzzle and conundrum undertaken for
the last 20 years suggest that there should be some alternative approach to their
resolution. In this respect our attention is attracted by graphene which is a 2D sheet
of carbon atoms in the form of a hexagonal crystal lattice. The outstanding property
of graphene is that at energies below a few eV it is well described by the Dirac
model characterized by the linear dispersion relation where the speed of light $c$
is replaced with the Fermi velocity $v_F$ \cite{118}.
This makes it possible to calculate many graphene properties, including its dielectric
response to the electromagnetic field, on the basis of first principles of Quantum
Electrodynamics at nonzero temperature. In so doing the phenomenological
response functions given by the Drude or plasma models become unneeded.

Thus, it was found that graphene is described by the transverse and longitudinal
dielectric permittivities depending on both the frequency and the wave vector which
are expressed via the polarization tensor \cite{119,120,121,122}.
In doing so, the predictions of the Lifshitz theory using the exact graphene response
functions are in good agreement with an experiment on measuring the Casimir force
between an Au-coated sphere and a graphene-coated plate \cite{123,124}.
What is more, the Casimir and Casimir-Polder entropies calculated for both the
pristine graphene sheets and for graphene possessing the nonzero energy gap and chemical
potential satisfy the Nernst heat theorem \cite{125,126,127,128,129}.
This means that there is no Casimir puzzle for graphene.

The case of graphene suggests that the roots of the problems discussed above might
be in the inadequate response functions of conventional 3D materials used in the
Lifshitz theory. The point is that the Casimir free energy (\ref{eq3}) and force
(\ref{eq6}) are obtained by the integration ober all $\kb$ from zero to infinity
at each fixed $\omega$. This means that both the propagating waves, which are on the
mass shell, and evanescent waves, which are off the mass shell, contribute to the
final result. The Drude response function is well checked experimentally and
provides the correct response to real electromagnetic fields with a nonzero field
strength. There is no direct experimental confirmation to this model, however,
if we are seeking for the response to quantum fluctuations which are off the mass
shell. What's more, the Casimir puzzle can be considered as an indirect indication
that the Drude model describes the response to fluctuations of such kind incorrectly.

At this point it is reasonable to search for the nonlocal Drude-like response
functions which provide an approximately the same response, as the standard Drude
model, to the electromagnetic fluctuations and fields on the mass shell but describe
adequately the response to the off-shell fluctuations. The response functions of
this kind were recently suggested in \cite{130}. They are given by
\begin{eqnarray}
&&
\tve_D^T(\omega,\kb)=1-\frac{\omega_p^2}{\omega[\omega+{\rm i}\gamma(T)]}
\left(1+{\rm i}\frac{v^T\kb}{\omega}\right),
\nonumber \\
&&
\tve_D^L(\omega,\kb)=1-\frac{\omega_p^2}{\omega[\omega+{\rm i}\gamma(T)]}
\left(1+{\rm i}\frac{v^L\kb}{\omega}\right)^{-1},
\label{eq60}
\end{eqnarray}
\noindent
where the quantities $v^{T,L}$ are the constants of the order of the Fermi velocity
$v_F$. It is seen that in the local limit both $\tve_D^T$ and $\tve_D^L$ coincide with
the standard Drude model
\begin{equation}
\tve_D^T(\omega,0)=\tve_D^L(\omega,0)=\ve_D(\omega).
\label{eq61}
\end{equation}

For the electromagnetic fields on the mass shell we have $c\kb\leqslant\omega$ and
\begin{equation}
\frac{v^{T,L}\kb}{\omega}\sim\frac{v_F}{c}\,\frac{c\kb}{\omega}
\leqslant\frac{v_F}{c}\ll 1.
\label{eq62}
\end{equation}
\noindent
Thus, the dielectric functions (\ref{eq60}) lead in this case to approximately
the same results as the standard Drude model. This is quite different from the
nonlocal response functions discussed in Section~7.6. For example, the nonlocal response
describing the anomalous skin effect \cite{109} used in theory of the Casimir force
\cite{112,113} (see Section 7.6)  was introduced for a description of the
physical phenomenon occurring in the fields on the mass shell. The nonlocal response
functions describing the electron gas without and with account of collisions were
introduced in \cite{131,132}. In the static limit they describe the screening effects
considered in Section~7.7. These again occur in real fields on the mass shell.

By contrast, the suggested permittivities (\ref{eq60}) are not intended for a
description of small nonlocal effects in real fields and deviate from the standard
Drude model only for the off-the-mass-shell fluctuations. Note also that the
permittivities (\ref{eq60}) depend only on $\kb$ and do not depend on $k_z$ as in the
case of \cite{109,131,132}. The reason is that in the presence of the Casimir plates
the translational  invariance along the $z$ axis is violated and it is strictly
speaking impossible to define the response functions depending on the 3D vector
{\boldmath$k$} \cite{133,134}.

For the permittivities depending only on $\kb$ (\ref{eq55}) results in
\begin{eqnarray}
&&
\hspace*{-5mm}
Z_{\rm TM}(\ix_l,\kb)=\frac{c}{\xi_l}\left[\frac{\kb}{\veL}+
\frac{k^T(\ix_l,\kb)-\kb}{\veT}\right],
\nonumber \\
&&
Z_{\rm TE}(\ix_l,\kb)=\frac{\xi_l}{c k^T(\ix_l,\kb)}.
\label{eq63}
\end{eqnarray}
\noindent
Then the reflection coefficients (\ref{eq54}) valid in the nonlocal case are given by
$$r_{\rm TM}(\ix_l,\kb)=\frac{\veT q_l- k^T(\ix_l,\kb)-\kb\left[\veT-\veL\right]
\left[\veL\right]^{-1}}{\veT q_l+k^T(\ix_l,\kb)+\kb\left[\veT-\veL\right]
\left[\veL\right]^{-1}},$$
\begin{equation}
r_{\rm TE}(\ix_l,\kb)=\frac{q_l-k^T(\ix_l,\kb)}{q_l+k^T(\ix_l,\kb)}.
\label{eq64}
\end{equation}

Now it is possible to substitute the specific Drude-like permittivities (\ref{eq60})
taken at $\omega=\ix_l$ in  (\ref{eq64}), calculate the Casimir force between two Au plates
(\ref{eq9}) and, by using the PFA (\ref{eq33}), find the gradient of a Casimir force between a
sphere and a plate. After introducing small corrections due to surface roughness and
inaccuracies in the PFA, considered in Sections~7.3 and 7.4, one can compare the
theoretical results obtained using the Lifshitz theory with the nonlocal Drude-like
response functions (\ref{eq60}) with the measurement data.

\begin{figure}[!h]
\centering
\vspace*{-2.2cm}
\hspace*{-1.6cm}\includegraphics[width=16 cm]{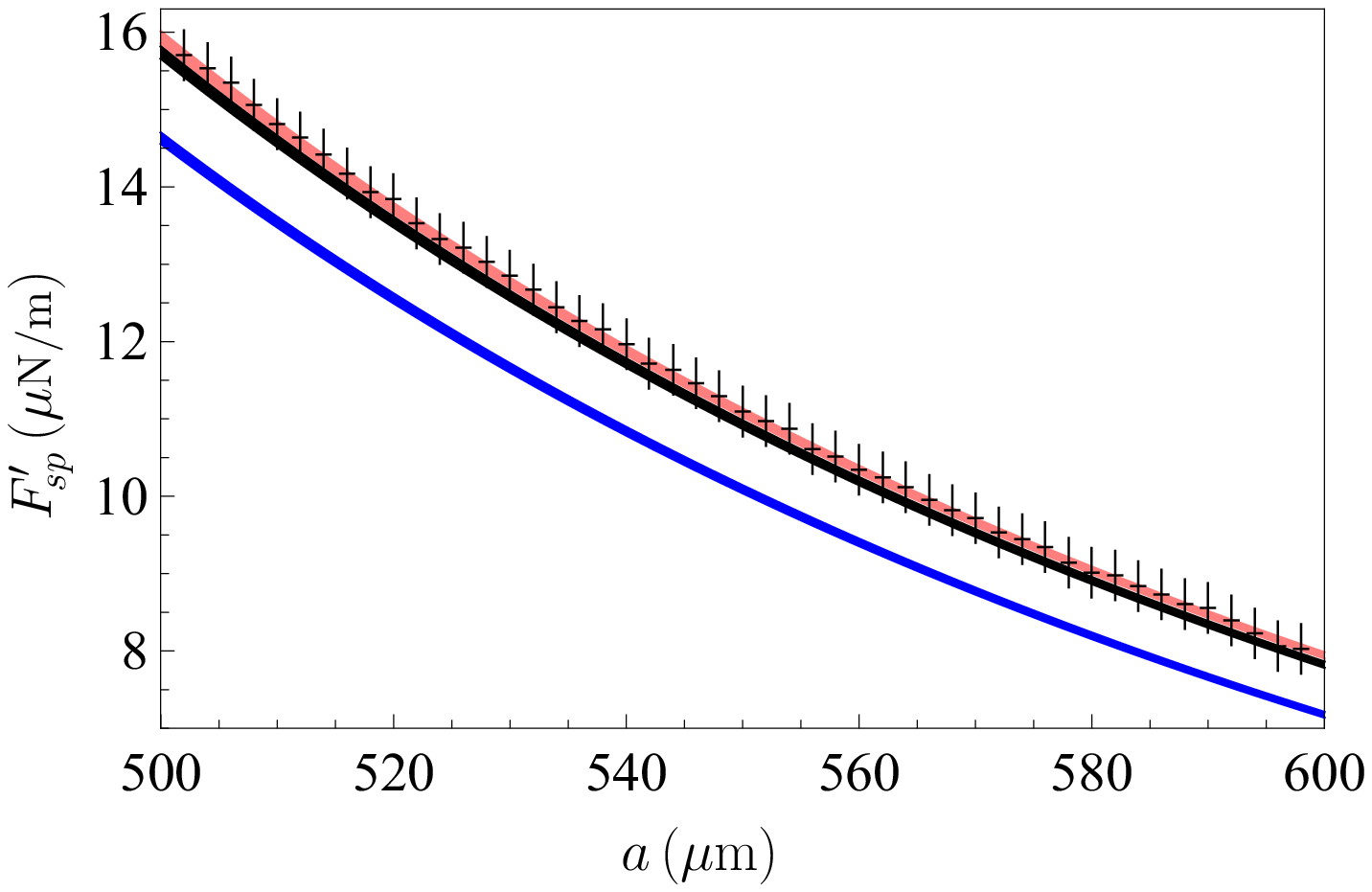}
\vspace*{-13.cm}
\caption{The gradients of the Casimir force between an Au-coated sphere and
an Au-coated plate measured by means of micromechanical torsional oscillator
\cite{21} (crosses) are compared with theoretical predictions of the Lifshitz
theory using the extrapolation of the optical data of Au by means of nonlocal
Drude-like response functions (black band). The theoretical predictions using
the Drude and plasma extrapolations are shown by the blue and red bands,
respectively.
\label{fg8}}
\end{figure}

In Figure~\ref{fg8} the obtained theoretical predictions in the experimental
configuration of the micromechanical torsional oscillator \cite{21} are shown as
a function of separation by the black band. The measurement results are shown as
crosses (the same as in Figure~\ref{fg2}). We also reproduce from Figure~\ref{fg2}
the red and blue theoretical bands found using the extrapolations of the optical
data by means of the plasma and Drude models. As is seen from Figure~\ref{fg8},
the theoretical predictions using both the nonlocal Drude-like and plasma response
functions are in agreement with the measurement data. An important advantage of
the Drude-like response is that it takes the proper account of the relaxation
properties of free charge carriers which are disregarded by the plasma model.

\begin{figure}[!t]
\centering
\vspace*{-2.2cm}
\hspace*{-1.6cm}\includegraphics[width=16 cm]{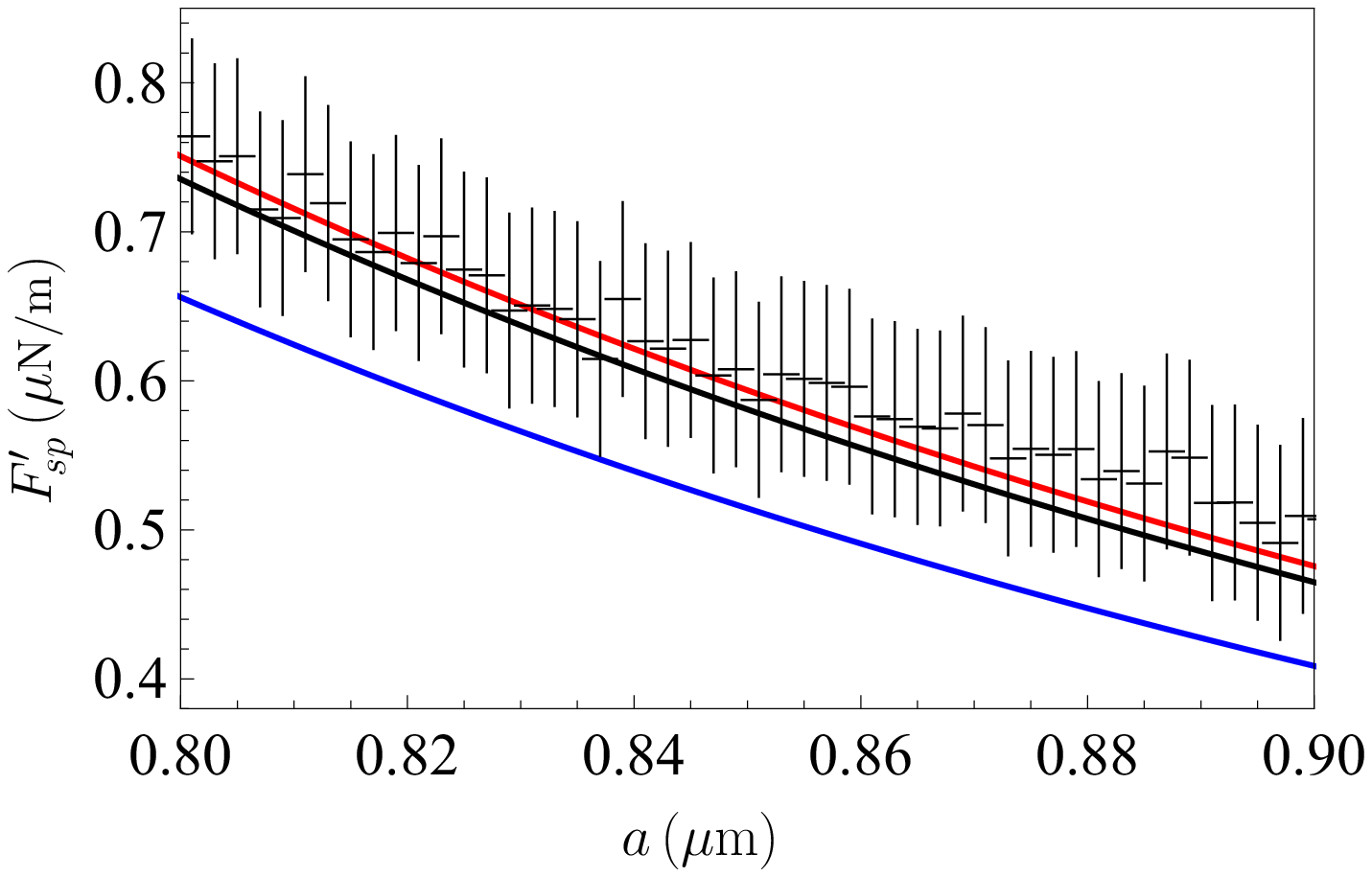}
\vspace*{-13.cm}
\caption{The gradients of the Casimir force between an Au-coated sphere and
an Au-coated plate measured by means of an atomic force microscope
\cite{78} (crosses) are compared with theoretical predictions of the Lifshitz
theory using the extrapolation of the optical data of Au by means of nonlocal
Drude-like response functions (black band). The theoretical predictions using
the Drude and plasma extrapolations are shown by the blue and red bands,
respectively.
\label{fg9}}
\end{figure}

As one more example, in Figure~\ref{fg9} we plot by the black band the predictions
of the Lifshitz theory using the optical data of Au extrapolated by the nonlocal
Drude-like response functions (\ref{eq60}) in the experiment measuring the gradient
of the Casimir force between an Au-coated sphere and an Au-coated plate at larger
separations by means of dynamic atomic force microscope \cite{78}
(see Section~7.2). The theoretical predictions using the Drude and plasma models
are shown as the blue and red bands, respectively, whereas the measured gradients
of the Casimir force are shown as crosses. It is again seen that the theoretical
approach using the nonlocal Drude-like response functions is in agreement with
the measurement data along with the plasma model. The advantages of the Drude-like
functions are that they take the proper account of dissipation of conduction electrons
and describe correctly the reflectances of the electromagnetic waves on the mass shell
incident on an Au plate \cite{130}.

Based on the above one can conclude that the nonlocal Drude-like response functions
(\ref{eq60}) provide a resolution of the experimental part of the Casimir puzzle for
nonmagnetic metals. Detailed examination of the asymptotic behavior of the Casimir
entropy calculated using the Drude-like functions (\ref{eq60}) at low temperature
demonstrates that these response functions also solve the thermodynamic part of the
Casimir puzzle. Thus, for nonmagnetic metals with perfect crystal lattices the
low-temperature behavior of the Casimir entropy is given by
\begin{equation}
\tilde{S}_D(a,T)=C_1(a)\sqrt{T},
\label{eq65}
\end{equation}
\noindent
where the coefficient $C_1$, in addition to separation, depends on $\omega_p$,
$v^T$ and the fundamental constants $c$, $\hbar$, and $k_B$.

For metal with impurities possessing some residual relaxation $\gamma_0$ the
low-temperature behavior of the Casimir entropy found using the nonlocal Drude-like
response functions (\ref{eq60}) takes the form
\begin{equation}
\tilde{S}_D(a,T)=C_2(a){T},
\label{eq66}
\end{equation}
\noindent
where, in addition to the parameters indicated above, the coefficient $C_2$ also
depends on $\gamma_0$.

{}From (\ref{eq65}) and (\ref{eq66}) it is seen that both for metals with perfect
crystal lattices and for metals with impurities the Casimir entropy goes to zero
with vanishing temperature. This means that the Lifshitz theory using the nonlocal
Drude-like response functions is consistent with the laws of thermodynamics.

\section{Discussion: The Present Status of the Casimir Puzzle and Casimir Conundrum}

In the foregoing, we have considered several theoretical and
experimental results which are not fully understood up to the
present time and were called in the literature the Casimir puzzle
and the Casimir conundrum. These results are related to the Casimir
effect which is the physical phenomenon determined by the
properties of the quantum vacuum and its interaction with matter.
As discussed in Section 1, several basic challenges of modern
physics are directly connected with the concept of vacuum. Because
of this, it is probably not surprising that the Casimir puzzle and
the Casimir conundrum have no an ultimate resolution until now in
spite of much work made by many researches.

Several precise experiments on measuring the Casimir interaction
between metallic and dielectric test bodies confirmed the fact
that the standard dielectric functions commonly used for a
description of the response of matter to real electromagnetic
fields with a nonzero field strength lead to incorrect predictions
in the framework of the Lifshitz theory. Over a period of time,
there were serious doubts regarding the precision and interpretation
of these measurements but after the experiments with magnetic test
bodies \cite{24,25} and differential force measurement, where the
measured signal differed from theoretical prediction of the Lifshitz
theory by the factor of 1000 (see Section 4.2), the facts have been
established with certainty.

Taking into account that the Drude response function also has an
unambiguous experimental confirmation in the area of real
electromagnetic fields with a nonzero field strength, it is
reasonable to suggest that it may describe incorrectly the response
of metals to quantum fluctuations off the mass shell. This suggestion
found support in the fact that some indirect experimental evidence
can be obtained only about the form of the longitudinal permittivity
$\tilde{\ve}^L$ in the off the mass shell fields, but not about
$\tilde{\ve}^T$ \cite{108} which should be derived theoretically
like this is done for graphene using the polarization tensor.

No less difficulties are connected with the thermodynamic parts of
the Casimir puzzle and the Casimir conundrum. Contradictions between
the Lifshitz theory and thermodynamics arise when we use the response
functions excluded by measurements of the Casimir force. Thus, it is
reasonable to hope that an agreement will be restored if the used
response to the off the mass shell quantum fluctuations, which is
extrapolated from the area of real fields with a nonzero field strength
and has no direct experimental confirmation, will be somehow corrected.

In the case of nonmagnetic metals possible realization of this program
has already been suggested (see Section 8) in the form of nonlocal
Drude-like response functions which lead to approximately the same
results, as the standard Drude model, in the on the mass shell fields,
but to significantly different response to the off-the-mass-shell
quantum fluctuations. It was shown that the Lifshitz theory using the
nonlocal Drude-like model is in agreement with the measurement data,
as it does when using the plasma model which simply disregards the
relaxation properties of conduction electrons. What is more, the
Casimir entropy calculated using the Drude-like response functions
satisfies the Nernst heat theorem for metals with both perfect
crystal lattices and lattices with impurities. Thus, the suggested
Drude-like model points the way to a resolution of the Casimir
puzzle.

We emphasize, however, that, unlike the case of graphene, this is a
phenomenological solution yet. It demonstrates the possibility to
solve the Casimir puzzle by modifying the dielectric response to
quantum fluctuations off the mass shell, but this solution might be
not unique. Furthermore, it is necessary to extend the proposed
approach to the case of magnetic metals and apply similar ideas
to dielectric materials in order to find a resolution of the
Casimir conundrum. Thus, further investigations of these challenging
problems are necessary.

\section{Conclusions}

To conclude, we have elucidated current situation regarding the
complicated problems faced by the Lifshitz theory during the last 20
years. It is shown that the experimental facts in this field of
research are largely settled. However, their theoretical understanding
is far from being complete. There are some ideas and even examples of
considerable promise on how the Casimir puzzle and the Casimir
conundrum could be solved but they need further theoretical
justification.

The thermodynamic test of the Lifshitz theory using one or other
type of the dielectric response confirmed its usefulness. It is
not by chance that the Lifshitz theory using the Drude model for
metals or the dc conductivity for dielectrics was found to violate
the Nernst heat theorem. A development of the Casimir physics
confirmed that the thermodynamic and experimental problems happen
concurrently. This means that if the Lifshitz theory using some model
of the dielectric response leads to a violation of the Nernst heat
theorem for the Casimir entropy one should expect that theoretical
predictions using this model will be found in disagreement with the
measurement data.

The suggested nonlocal Drude-like response functions provide a
supposed resolution for both the experimental and thermodynamic
parts of the Casimir puzzle for nonmagnetic metals. Future trends
should bring the final resolution of the complicated problems of
Casimir physics for both metallic and dielectric materials and
widen the scope of fluctuational electrodynamics by a more
reliable description of the dielectric response of matter to
quantum fluctuations off the mass shell.

\funding{This work was supported by the Peter the Great Saint Petersburg Polytechnic
University in the framework of the Russian state assignment for basic research
(project N FSEG-2020-0024).
The work was also partially funded by the Russian Foundation for Basic Research grant number
19-02-00453 A. The author was partially supported by the Russian Government Program of Competitive
Growth of Kazan Federal University. }

\acknowledgments{
The author is grateful to A.A. Banishev, V.B. Bezerra,
E.V. Blagov, M. Bordag, R. Castillo-Garza, C.-C. Chang, F. Chen, R.S. Decca, E. Fischbach,
B. Geyer, R.K. Kawakami, G.L. Klimchitskaya, C.C. Korikov, D.E. Krause, M. Liu,
D. L\'{o}pez, U. Mohideen, V.M. Petrov, C. Romero, H. Wen, J. Hu for collaboration work
on our joint articles cited in this review.}

\conflictsofinterest{The author declares no conflict of interest.}

\reftitle{References}

\end{document}